\documentclass[11pt]{article}
\newlength{\vshift}
\newlength{\hshift}
\usepackage{amssymb}
\usepackage{amsmath}
\usepackage{amsthm}
\usepackage{amsopn}
\def\diff{\textrm{d} }
\def\nn{\nonumber }
\def\la{\lambda}

\def\ka{\kappa}
\def\de{\delta}
\def\be{\beta}

\def\al{\alpha}
\def\si{\sigma}

\def\shouldid{\stackrel{!}{=}}

\def\ds{\stackrel{\star}{,}}

\def\hxi{\hat \xi}
\def\hdi{\hat D}
\def\hom{\hat \omega}
\def\td{\textrm{d}}

\def\xx{{\hat x}}

\def\hp{\hat{\partial}}
\def\p{\partial}

\def\hp{\hat{\partial}}
\def\h{\hat}
\def\lb{\lbrack}
\def\rb{\rbrack}
\def\pat{\partial}

\providecommand{\href}[2]{#2}

\begin{document}

\begin{titlepage}
\rightline{LMU-TPW 2004-03}
\rightline{MPP-2004-44}

\vspace{4em}
\begin{center}

{\Large{\bf Derivatives, forms and vector fields\\ on the
    $\kappa$-deformed Euclidean space}}

\vskip 3em

{{\bf Marija Dimitrijevi\' c${}^{1,2,3}$,
    Lutz M\"oller${}^{1,2}$, Efrossini Tsouchnika${}^{1}$}}

\vskip 1em

${}^{1}$Universit\"at M\"unchen, Fakult\"at f\"ur Physik\\
        Theresienstr.\ 37, D-80333 M\"unchen\\[1em]

${}^{2}$Max-Planck-Institut f\"ur Physik\\
        F\"ohringer Ring 6, D-80805 M\"unchen\\[1em]

${}^{3}$University of Belgrade, Faculty of Physics\\
Studentski trg 12, SR-11000 Beograd\\[1em]
 \end{center}

\vspace{2em}

\begin{abstract}
The model of $\kappa$-deformed space is an interesting
example of a noncommutative space, since it allows
a deformed symmetry.

In this paper we present new results concerning
different sets of derivatives on the coordinate algebra of 
$\kappa$-deformed Euclidean space. We introduce a differential
calculus with two interesting sets of one-forms and higher-order
forms. The transformation law of vector fields is constructed in
accordance with the 
transformation behaviour of  derivatives. The crucial property of
the different derivatives, forms and vector fields is that in an $n$-dimensional
spacetime there are always $n$ of them. This is the key difference with respect to conventional
approaches, in which the differential calculus is ($n+1$)-dimensional.

This work shows  that
derivative-valued quantities such as derivative-valued vector fields
appear in a generic way on noncommutative spaces.

\end{abstract}
\vskip 0.5cm
\qquad\hspace{2mm}\scriptsize{eMail: dmarija;  lmoeller;
frosso@theorie.physik.uni-muenchen.de}
\vfill

\end{titlepage}\vskip.2cm

\newpage
\setcounter{page}{1}

\section{Introduction}
\label{A}

The model of $\kappa$-deformed spacetime has originally been
introduced in \cite{lukrue1}, \cite{lukrue2} (for a more comprehensive list of references,
see \cite{f1})  and has afterwards been discussed by several groups,
both from a mathematical and a physical perspective.

The main interest in the  $\kappa$-deformed spacetime comes from the
fact that this model is a
mild deformation of spacetime. There
is only one coordinate that does not commute with all
others. Therefore this is a sufficiently
simple model that may serve as a playground to develop generic
concepts for noncommutative spaces. The reason is that it is  a mathematically consistent deformation of
spacetime, compatible with a simultaneous
deformation of the symmetry structure, the $\kappa$-Poincar\'e
algebra.  Recently, a new motivation for
studying the  $\kappa$-deformed
spacetime has appeared. Namely, it is a  well-founded framework for so-called doubly special
relativity, i.e. special relativity with a second invariant scale
(cp. \cite{gac1}, \cite{luk2004}).

In its Minkowski version the $\kappa$-deformed spacetime has first been treated as the
translational sector of the $\kappa$-Poincar\'e group. It contains one
distinguished coordinate which does
not commute with all other coordinates. The
$\kappa$-Poincar\'e group as a Hopf algebra is dual to the
$\kappa$-Poincar\'e algebra \cite{kosmas}, which  has first been derived contracting  the Hopf algebra
$SO_q(3,2)$. It has been found that in the so called bicrossproduct basis
\cite{majrue}   the generators of the
$\kappa$-Lorentz algebra fulfil the commutation relations of 
 the undeformed Lorentz Lie algebra. Nevertheless, the symmetry
 generators of the 
$\kappa$-Poincar\'e  Hopf algebra
act in a deformed way on products of functions.

The
bicrossproduct basis of the $\kappa$-Poincar\'e algebra can be
obtained by a constructive procedure requiring consistency with the
algebra of coordinates. The transformation behaviour of the
coordinates under rotations leads to an undeformed algebra of
rotations.  Consistency with this undeformed algebra of rotations
and consistency with the algebra of
 coordinates are then the two touchstones to construct geometrical
 concepts such as derivatives, forms and vector fields. This
 is the content of this paper. It complements
several analyses concerning the $\kappa$-deformed spacetime which we
recently presented in \cite{f1}, \cite{f2}. Our construction of a
field theory based on representing quantities of $\ka$-deformed spacetime
by means of $\star$-products is similar to other recent approaches \cite{KLM}
and \cite{gac2}.

This work is organised as follows: in the subsequent section we fix
our notation concerning the abstract algebra, recapitulating  results of
\cite{f1}. 
We present the Hopf algebra
properties of our model  in an introductory way. Several
sets of derivatives are discussed; they  either have simple commutation relations
with the coordinates or simple transformation behaviour under rotations.

In section three we  discuss  three $\star$-products which can be
defined in a generic way. A closed formula for the symmetric
$\star$-product is derived. The symmetry generators are represented in
terms of  derivative operators both for the symmetric
$\star$-product and for normal ordered $\star$-products.

In section four we present an analysis of the differential calculus on
the $\kappa$-deformed space. In contrast to results in the literature we argue
that an  $n$-dimensional $\kappa$-deformed space can be equipped with an $n$-dimensional
differential calculus of one-forms. To be able to do so, we have to
accept that the
commutation relations of one-forms with coordinates become
derivative-valued. We calculate
frame
one-forms, which commute with the coordinates, and construct
representations of the
forms, regarding them
as derivative-valued maps in the algebra of functions of commuting variables.

In section five we introduce vector fields
 by generalising the transformation behaviour of
derivatives under the $\kappa$-deformed rotations. We construct maps
between different vector fields and
find that generically also vector fields are derivative-valued quantities.

%
%
%
%
%
%
%
%
%
%
%
%

\section{Derivatives on the $\ka$-deformed space}
\subsection{The  $\ka$-deformed space}
\label{B}
The $n$-dimensional $\ka$-deformed space is a noncommutative space of the Lie
algebra type, i.e. it is the associative factor space algebra $\mathcal{A}_\xx$ freely generated
by $n$ abstract  coordinates $\xx^\mu$, divided by the ideal freely
generated by the commutation relations \cite{madore}
\begin{equation}
\label{k4}
\lb \hat{x}^\mu, \hat{x}^\nu\rb = iC^{\mu\nu}_\la \hat{x}^\la.
\end{equation}
The structure constants for the $\ka$-deformed space are
 $C^{\mu\nu}_\la=a^\mu\de^\nu_\la-a^\nu\de^\mu_\la$. The
 characteristic deformation vector  $a^\mu$ can be rotated without loss of
 generality into one of the $n$ directions 
 $a^\mu=a\de^\mu_n$. In this case there is
 one special coordinate $\hat{x}^n$,
which does not commute with all other coordinates
\begin{equation}
\label{k1}
[\hat{x}^n,\hat{x}^j]=ia\hat{x}^j,\qquad [\hat{x}^i,\hat{x}^j]=0, \qquad i,j=1,2,\dots,n-1.
\end{equation}
Coordinates without
hat such as $x^\mu$ denote ordinary variables. In most discussions of $\ka$-deformed spacetime, $\hat{x}^n$ is
taken to be the time coordinate of a four-dimensional Minkowski
spacetime. The restriction to four dimensions had already been lifted in
\cite{lukruegg}. Identifying $\hat{x}^n$
with the direction of time is an additional choice. In our approach
$\hat{x}^n$ is an arbitrary
direction of an $n$-dimensional Euclidean space. This setting is chosen for transparency of the calculus.
Indices can arbitrarily be lowered or raised with
a formal metric $g^{\mu\nu}=\de^{\mu\nu}$. Greek indices take values
$1,\dots,n$, while  Roman indices apart
from $n$ run from $1,\dots,n-1$.
We use the Einstein summation
conventions.

Our formulae generalise in a straightforward way to a
Minkowski setting \cite{f2}. Also for non-diagonal metrics
 there are generalisations \cite{klm0307}.  We use the notation (\ref{k1}) instead
of  $\kappa=\frac{1}{a}$, since this is more convenient for
working  in configuration space.

Derivatives are regarded as maps in the coordinate algebra,
$\widehat{\pat}_\mu:\mathcal{A}_\xx\rightarrow\mathcal{A}_\xx$. In order to define derivatives $\widehat{\partial}_\mu$, we
demand that they
\begin{itemize}
\item are consistent with
(\ref{k1});
\item  are a
deformation of ordinary derivatives, i.e.
$\lb\widehat{\partial}_\mu,\hat{x}^\nu\rb=\de_\mu^\nu
+\mathcal{O}(a)$;
\item  commute among themselves.
\end{itemize}
These restrictions on derivatives $\widehat{\partial}_\mu$
are weak, there exists a wide range of possible solutions
\begin{equation}
\label{k2a}
\lb\widehat{\partial}_\mu,\hat{x}^\nu\rb=\de_\mu^\nu + \sum_{j}a^j (\widehat{\partial}_{(\mu,\nu)})^j.
\end{equation}
The symbolic notation denotes all terms of a power series expansion
in the derivatives $\widehat{\partial}_\mu$, which are consistent
with the index structure.

With the additional condition that the commutator
$\lb\widehat{\partial}_\mu,\hat{x}^\nu\rb$ is linear in the
derivatives, there are three one-parameter families of solutions:
\begin{eqnarray}
\label{k3}
\lb \hat{\partial}^{c_1}_n, \hat{x}^n\rb &=& 1+iac_1\hat{\partial}^{c_1}_n,\quad\hspace{0.5mm}
\lb \hat{\partial}^{c_2}_n, \hat{x}^n\rb = 1+iac_2 \hat{\partial}^{c_2}_n,\qquad\hspace{4mm}
\lb \hat{\partial}^{c_3}_n, \hat{x}^n\rb = 1+2ia \hat{\partial}^{c_3}_n,\nonumber \\
\lb \hat{\partial}^{c_1}_n, \hat{x}^j\rb &=&0,\qquad\qquad\hspace{5.5mm}
\lb \hat{\partial}^{c_2}_n, \hat{x}^j\rb =ia(1+c_2)\hat{\partial}^{c_2}_j,\qquad\hspace{1mm}
\lb \hat{\partial}^{c_3}_n, \hat{x}^j\rb =ia\hp^{c_3}_j,\nonumber \\
\lb \hat{\partial}^{c_1}_i, \hat{x}^n\rb &=&ia \hat{\partial}^{c_1}_i,\qquad\quad\hspace{3mm}
\lb \hat{\partial}^{c_2}_i, \hat{x}^n\rb =0,\qquad\qquad\qquad\hspace{4.5mm}
\lb \hat{\partial}^{c_3}_i, \hat{x}^n\rb =iac_3\hp_i^{c_3},  \\
\lb \hat{\partial}^{c_1}_i, \hat{x}^j\rb &=& \delta_i^j, \qquad\qquad\hspace{4.2mm}
\lb \hat{\partial}^{c_2}_i, \hat{x}^j\rb = \delta_i^j(1+iac_2 \hat{\partial}^{c_2}_n ), \quad\hspace{1.5mm}
\lb \hat{\partial}^{c_3}_i, \hat{x}^j\rb = \delta_i^j. \nonumber
\end{eqnarray}
The real parameters $c_i$ are not fixed by
consistency with (\ref{k1}). We prefer to work with one particular
choice in the following, $\hp_\mu^{c_1=0}$. For brevity,
 $\hp_\mu^{c_1=0}$ is denoted as $\hp_\mu$.
There is always more than one set of linear derivatives (consistent with the coordinate algebra) on noncommutative spaces of the Lie
algebra type (\ref{k4}).
If we denote the commutator of coordinates and derivatives linear in
$\widehat{\partial}_\mu$ as
\begin{equation}
\label{k5}
\lb \widehat{\partial}_\mu, \hat{x}^\nu\rb = \de_\mu^\nu+ i\rho_\mu^{\nu\la}\widehat{\partial}_\la, 
\end{equation}
we obtain two conditions on $\rho_\mu^{\nu\la}$ from consistency with (\ref{k4}):
\begin{equation}
\label{k6}
\rho^{\mu\nu}_\la-\rho^{\nu\mu}_\la=C^{\mu\nu}_\la, \qquad
\rho^{\mu\nu}_\la \rho^{\ka\si}_\nu - \rho^{\ka\nu}_\la
\rho^{\mu\si}_\nu = C^{\mu\ka}_\nu\rho^{\nu\si}_\la.
\end{equation}
 All three one-parameter sets of derivatives $\hp_\mu^{c_i}$ (\ref{k3})
fulfil the conditions (\ref{k6}). With the freedom 
indicated by the parametrisation in (\ref{k3}), we have exhausted 
all linear derivatives. 
That there is such a variety of  linear derivatives is disturbing at first sight. But all
three families $\hp_\mu^{c_i}$ can be mapped into each other.
The derivatives $\hp_\mu$ ($c_1=0$) are mapped to the derivatives
$\hp_\mu^{c_1}$ for arbitrary $c_1$ in the following way:
 \begin{equation}
\label{k6d}
\hp_j^{c_1}=\hp_j,\qquad\hspace{5mm} \hp_n^{c_1}=\frac{e^{iac_1\hp_n}-1}{iac_1}.
\end{equation}
The role of shift operators is played by the following operators, in terms of $\hp^{c_1}_\mu$ 
\begin{equation}
\label{k6g}
e^{-ia\hp_n}=\big(\frac{1}{1+ic_1a\hp^{c_1}_n}\big)^{\frac{1}{c_1}},\hspace{15mm}e^{ia\hp_n}= (1+ic_1a\hp^{c_1}_n)^{\frac{1}{c_1}}.
\end{equation}

The derivatives
$\hp^{c_2}_\mu$  can be expressed in terms of
$\hat{\p}_\mu$ as well:
\begin{equation}
\label{k6f}
\hp^{c_2}_n = \frac{e^{iac_2\hp_n}-1}{iac_2} ,\qquad
\hp^{c_2}_j = \hp_j e^{iac_2\hp_n},\quad e^{ia\hp_n}=(1+ic_2a\hp^{c_2}_n)^{\frac{1}{c_2}}.
\end{equation}
The map from $\hp_\mu$ to derivatives $\hp_\mu^{c_3}$ reads
\begin{equation}
\label{k6r}
\hp^{c_3}_n =
\frac{e^{2ia\hp_n}-1}{2ia}+\frac{iac_3}{2}\hp_k\hp_k,\qquad\hp^{c_3}_j
=\hp_j,\qquad e^{ia\hp_n}=(1+2ia\hp^{c_3}_n +a^2c_3\hp^{c_3}_l\hp^{c_3}_l)^{\frac{1}{2}}.
\end{equation}

The derivatives $\hp_\mu$ are a very suitable
basis in the algebra of derivatives to develop our
formalism, as we will see later. The maps
(\ref{k6d}), (\ref{k6f}) and (\ref{k6r}) allow to reformulate the
entire formalism, which we will develop in the
following in terms of $\hp_\mu$, also in terms of any of the three one-parameter families of linear
derivatives $\hp_\mu^{c_i}$. 

We will discuss
several other derivatives  in the following, for which we
will lift the condition that the commutator (\ref{k2a}) is linear in $\widehat{\partial}_\mu$.

%
%

\subsection{$\mathbf{SO_a(n)}$ as a Hopf algebra}

We define the generators of rotations $ M^{\mu\nu}$  by their
commutation relations with the coordinates, demanding consistency with
 (\ref{k1}). We require undeformed transformation behaviour, i.e. the
commutation relations of the Lie algebra of $SO(n)$,  to zeroth
order. In addition the generators of rotations $M^{rs}$ and
$N^l=M^{nl}$ should appear at most linearly on the right hand side of the commutators  $\lb
 M^{\mu\nu},\hat{x}^\rho\rb $. The only terms admissible in
 $\mathcal{O}(a)$ therefore involve the generators of rotations $M^{rs}$ and
$N^l=M^{nl}$ exactly once. Higher order terms  in $a$
have to be accompanied by derivatives because of dimensional
reasons. If we demand\footnote{More general commutation relations follow if this
  requirement is lifted.} that the commutation relations close in
coordinates and generators of rotations alone, the unique solution
consistent with (\ref{k1}) is:
\begin{eqnarray}
\label{k7}
\lb M^{rs}, \hat{x}^n\rb &=& 0,\hspace{30mm}  \lb N^{l}, \hat{x}^n\rb = \hat{x}^l+ia N^{l} ,\nonumber \\
\lb M^{rs}, \hat{x}^j\rb &=& \de^{rj}\hat{x}^s -\de^{sj}\hat{x}^r,\hspace{10mm}
\lb N^{l}, \hat{x}^j\rb =  - \de^{lj}\hat{x}^n -ia M^{lj}.
\end{eqnarray}
The generators $M^{\mu\nu}$ have
the commutation relations of the Lie algebra of $SO(n)$:
\begin{eqnarray}
\label{k7a}
\lb M^{rs}, M^{tu}\rb &=& \de^{rt}M^{su}+\de^{su}M^{rt}-\de^{ru} M^{st}-\de^{st}M^{ru},\nonumber \\
\lb M^{rs}, N^i\rb &=& \de^{ri} N^s-\de^{si} N^r,\hspace{10mm}
\lb N^{i}, N^j\rb = M^{ij}.
\end{eqnarray}
Even if, according to (\ref{k7a}), the algebra of rotations is undeformed, the action
on the coordinates is deformed (\ref{k7}). Therefore we will call
$M^{\mu\nu}$ the generators of the algebra of $SO_a(n)$
rotations. Although actually the universal enveloping algebra of the Lie algebra
of the symmetry group is deformed, we use  the symbol of the
symmetry group with a slight abuse of notation.

We emphasise that consistency with the coordinate algebra
leads to the
so called bicrossproduct basis of the $\ka$-deformed Euclidean algebra,
first defined in \cite{majrue}. The bicrossproduct basis is singled
out by (\ref{k7a}) in contrast to the so called classical basis
which may be obtained by contracting  the  $q$-anti-de Sitter Hopf
algebra $SO_q(3,2)$ \cite{lukrue1}.
The classical and the bicrossproduct basis are related by a nonlinear
change of variables. We work in  the
bicrossproduct basis in the following. For all further constructions,
consistency with  (\ref{k1}) and (\ref{k7a}) is a crucial requirement.

An important ingredient of the symmetry structure of $\ka$-deformed
space are the Leibniz rules of the
 generators of rotations and the derivatives\footnote{We use the term ``Leibniz rule'' also for the
  action of the generators of rotations on products of functions.}. They  can be derived immediately from (\ref{k3})
and (\ref{k7}):
\begin{eqnarray}
\label{k9}
\hat{\partial}_n\hspace{1mm} (\hat{f} \cdot \hat{g}) & = & (\hat{\partial}_n \hat{f})
\cdot \hat{g} + \hat{f} \cdot (\hat{\partial}_n \hat{g}),\nonumber \\
\hat{\partial}_j\hspace{1mm} (\hat{f} \cdot \hat{g}) & = & (\hat{\partial}_j \hat{f})
\cdot \hat{g} + (e^{ia\hat{\partial}_n} \hat{f}) \cdot (\hat{\partial}_j \hat{g}),\nonumber \\
M^{rs}\hspace{1mm} (\hat{f} \cdot \hat{g}) & = & (M^{rs} \hat{f})\cdot \hat{g} + \hat{f}
\cdot (M^{rs} \hat{g}), \\
N^{l}\hspace{1mm} (\hat{f} \cdot \hat{g}) & = & (N^{l} \hat{f})\cdot \hat{g} +
(e^{ia\hat{\partial}_n} \hat{f})\cdot  (N^{l} \hat{g}) - ia (\hat{\partial}_j
\hat{f})\cdot (M^{lj} \hat{g})\hspace{1mm}.\nonumber
\end{eqnarray}

In a more technical language, equations (\ref{k9}) are the
coproducts:
\begin{eqnarray}
\label{k8}
\Delta \hat{\partial}_n & = & \hat{\partial}_n \otimes 1
 +  1\otimes \hat{\partial}_n,\nonumber \\
\Delta \hat{\partial}_j & = & \hat{\partial}_j\otimes  1
+ e^{ia\hat{\partial}_n} \otimes \hat{\partial}_j,\nonumber \\
\Delta M^{rs} & = & M^{rs}\otimes  1  +  1\otimes M^{rs}, \\
\Delta N^{l} & = & N^{l}\otimes 1  +
e^{ia\hat{\partial}_n}\otimes N^{l}  - ia \hat{\partial}_j
\otimes M^{lj} \hspace{1mm}.\nonumber
\end{eqnarray}

The notion of coproduct leads to the observation that the generators of the
$\ka$-deformed symmetry are elements of a Hopf algebra
$\mathcal{A}$. A Hopf algebra  is characterised by the specification
of five operations on  elements of a vector space: familiar operations
are multiplication of vector space elements, the product
($\mathcal{V}\cdot\mathcal{W}\in\mathcal{A}\hspace{2mm}$ if $\hspace{2mm}\mathcal{V},
\mathcal{W} \in \mathcal{A}$),  and the unit
$\mathbf{1}\in\mathcal{A}$
($\mathcal{V}\cdot\mathbf{1}=\mathbf{1}\cdot\mathcal{V}=\mathcal{V}$).
An \textit{algebra} with a unit  is a vector space which closes under
multiplication of its elements.

The concept of  \textit{coalgebra} is in an abstract sense dual to
the concept of an
algebra. For a coalgebra two operations on
vector space elements have to be specified: the coproduct $\Delta(\mathcal{V})$
and the counit $\epsilon(\mathcal{V})$.  In the language of representations, the
coproduct
specifies how a coalgebra element $\mathcal{V}\in\mathcal{A}$ acts on products of
representations. The counit describes the action on the zero-dimensional
representation.

For a \textit{bialgebra}, the  algebra aspects and the coalgebra aspects have to be
compatible.
For a \textit{Hopf algebra},  an additional operation, the antipode $S(\mathcal{V})$, has
to be defined, compatible with all other
operations. The antipode is the analog of the inverse element of
groups; in the language of representations, it states the action on
the dual representation.

In our case we obtain:
\begin{eqnarray}
\label{k10}
\epsilon(\hat{\partial}_n)=0,\quad && S(\hat{\partial}_n) = -\hat{\partial}_n,\nonumber\\
\epsilon(\hat{\partial}_j)=0,\quad&& S(\hat{\partial}_j)=-\hat{\partial}_j
e^{-ia\hat{\partial}_n},\\
 \epsilon(M^{rs})=0,\quad && S(M^{rs})=-M^{rs},\nonumber\\
\epsilon(N^l)=0,\quad && S(N^l)=-N^l
e^{-ia\hat{\partial}_n}-iaM^{lk}\hat{\partial}_k e^{-ia\hat{\partial}_n}-ia(n-1)\hat{\partial}_le^{-ia\hat{\partial}_n}.\nonumber
\end{eqnarray}

We have introduced  $M^{\mu\nu}$  in (\ref{k7}) as the generators of 
$SO_a(n)$ rotations. Since the coproduct involves derivatives, we can
 deform in a consistent way - as a Hopf algebra - only the (universal
 enveloping algebra of the) Lie algebra
 of the \textit{inhomogeneous} $SO(n)$. Under $SO_a(n)$ we will understand the deformed \textit{Euclidean Hopf
  algebra}. For most of the following issues, considering $SO_a(n)$ as a
bialgebra is sufficient. 

For groups the inverse of the inverse is the identity and the dual
representation of the dual representation is again the original
one. Applying the antipode
twice, we see that this is not necessarily the case for a deformed Hopf algebra
such as $SO_a(n)$. We obtain
$S^2(\mathcal{V})= \mathcal{V}$ for $\mathcal{V}\ne N^i$ and for $N^i$
\begin{equation}
\label{k11}
S^2(N^l)= N^l+ia(n-1)\hat{\partial}_l\ne N^l.
\end{equation}

In (\ref{k3}) derivatives
$\hat{\partial}_\mu$ have been introduced as a minimal, linear deformation of commutative
partial derivatives. These derivatives $\hat{\partial}_\mu$ ($c=0$)
are a module of $SO_a(n)$, however, they have
complicated commutation relations with the generators of rotations
\begin{eqnarray}
\label{k12}
\lb M^{rs},\hat{\partial}_n\rb & = &0,\qquad
\lb M^{rs},\hat{\partial}_j\rb =  \de_j^r\hat{\partial}_s - \de_j^s\hat{\partial}_r, \nonumber \\
\lb N^{l},\hat{\partial}_n\rb & = &\hat{\partial}_i,\qquad
\lb N^{l}, \hat{\partial}_j\rb  =
\de^l_{j}\frac{1-e^{2ia\hat{\partial}_n}}{2ia}-\de_{j}^l\frac{ia}{2}\hat{\partial}_k\hat{\partial}_k +ia\hat{\partial}_l\hat{\partial}_j\hspace{1mm}.
\end{eqnarray}
The other linear derivatives (\ref{k3}) are modules of $SO_a(n)$ as
well.

We will demand, going beyond (\ref{k7a}), that the algebra sector of
the full deformed Euclidean Hopf algebra $SO_a(n)$ 
 should remain undeformed. Equations (\ref{k12}) therefore force us to introduce other
derivatives, which we will call Dirac derivatives, as generators of
translations in $SO_a(n)$. This will be the content of the following subsection.

As an aside we note the representation of the
orbital part of the  generators of $SO_a(n)$ rotations $M^{rs}$ and
$N^l$, in terms of  $\hat{x}^\mu$ and $\hat{\partial}_\mu$:
\begin{eqnarray}
\label{k13}
\hat{M}^{rs} &=& \hat{x}^s \hat{\partial}_r - \hat{x}^r
\hat{\partial}_s,\nonumber\\
\hat{N}^l &=& \hat{x}^l \frac{e^{2ia\hat{\partial}_n}-1}{2ia}
-\hat{x}^n \hat{\partial}_l +\frac{ia}{2}\hat{x}^l\hat{\partial}_k\hat{\partial}_k.
\end{eqnarray}
This representation can be derived from (\ref{k12}) and it is
consistent with
(\ref{k7}) and (\ref{k7a}).

%
%

\subsection{Invariants and Dirac operator}
A familiar  result \cite{majrue}  is that
  the lowest order  polynomial in the coordinates  invariant under
  $SO_a(n)$ rotations is not
  $\hat{x}^\mu \hat{x}^\mu$ but
\begin{equation}
\label{k14}
\hat{I}_1=\hat{x}^\mu \hat{x}^\mu-ia(n-1)\hat{x}^n.
\end{equation}
This polynomial is not invariant in the sense
$[M^{\mu\nu},\hat{I}_1]=0$, since
\begin{equation}
\label{k15}
\lb M^{rs}, \hat{I}_1\rb = 0,\hspace{15mm}\lb N^i, \hat{I}_1\rb = 2ia \hat{x}^\mu M^{\mu i} +a^2(n-2)N^i.
\end{equation}
The polynomial (\ref{k14}) can meaningfully be interpreted as an invariant,
since another invariant (in the sense of (\ref{k15})) is obtained if  we multiply it with any $SO_a(n)$-invariant expression from
the right.

Equation (\ref{k14}) is the lowest order  invariant in the coordinates
alone. The Laplace operator $\hat{\Box}$ is the lowest order invariant built out of
derivatives and it is truly invariant under $SO_a(n)$
rotations:
\begin{equation}
\label{k16}
\hat{\Box}=\hat{\partial}_k\hat{\partial}_k e^{-ia\hat{\partial}_n}+\frac{2}{a^2}(1-\cos(a\hat{\partial}_n)),\quad
 \textrm{with}\quad
 [N^i,\hat{\Box}]=0, \quad [M^{rs},\hat{\Box}]=0.
\end{equation}
All functions $\hat{\Box}\cdot f(a^2\hat{\Box})$ of the Laplace
operator are invariant as well and are
consistent with the classical limit $\hat{\Box}=\hp_\mu\hp_\mu + \mathcal{O}(a)$.

The Dirac operator $\hat{D}$ is defined as the invariant under
\begin{equation}
\label{k18}
[N^i,\hat{D}]+[n^i,\hat{D}] = 0, \qquad
[M^{rs},\hat{D}]+[m^{rs},\hat{D}]= 0, 
\end{equation}
where $n^i = \frac{1}{4}[\gamma^n,\gamma^i]$ and $m^{rs}=\frac{1}{4}[\gamma^s,\gamma^r]$ are the generators  of  
rotations for spinorial degrees of freedom,
with the Euclidean $\gamma$-matrices
$\{\gamma^\mu,\gamma^\nu\}=2\de^{\mu\nu}$. The components of the  Dirac operator
$\hat{D}= \gamma^\mu \hat{D}_\mu$ will be called Dirac
derivatives \cite{Dirac1}. They are derivatives transforming linearly under $SO_a(n)$ rotations: 
\begin{eqnarray}
\label{k20}
\lb N^i, \hat{D}_n\rb&=&  \hat{D}_i, \qquad\qquad  \lb N^i,\hat{D}_j\rb=-\de^{ij} \hat{D}_n, \nonumber \\
\lb M^{rs},\hat{D}_n\rb&=&0, \qquad \qquad \lb M^{rs},\hat{D}_j\rb=\de^{r}_{j}\hat{D}_s-\de^{s}_{j}\hat{D}_r.
\end{eqnarray}
There is a continuous range of solutions to (\ref{k20})
with  classical limit $\hdi_\mu = \hp_\mu+  \mathcal{O}(a)$:
\begin{eqnarray}
\label{k20a}
\hat{D}_n&=&
\Big(\frac{1}{a}\sin(a\hat{\partial}_n)+\frac{ia}{2}\hat{\partial}_k\hat{\partial}_k
e^{-ia\hat{\partial}_n}\Big)f(a^2\hat{\Box}),\nonumber\\
\hat{D}_j&=&\hat{\partial}_j e^{-ia\hat{\partial}_n}f(a^2\hat{\Box}).
\end{eqnarray}
The simplest solution of (\ref{k20}) is the one with $f=\mathbf{1}$. We choose
this solution to be \emph{the} Dirac derivative. It is a nonlinear derivative in the sense of (\ref{k2a}):
\begin{eqnarray}
\label{k22}
\lb\hat{D}_n, \hat{x}^i \rb&=& ia\hat{D}_i, \nonumber \\
\lb\hat{D}_n, \hat{x}^n \rb&=&
\sqrt{1-a^2\hat{D}_\mu\hat{D}_\mu}= 1-\frac{a^2}{2}\hat{\Box}, \nonumber\\
\lb\hat{D}_j, \hat{x}^i \rb&=& \de^i_j\Big(-ia\hat{D}_n+
\sqrt{1-a^2\hat{D}_\mu\hat{D}_\mu}\Big)=\de^i_j\Big(1-ia\hat{D}_n-\frac{a^2}{2}\hat{\Box}
\Big), \\
\lb\hat{D}_j, \hat{x}^n \rb&=& 0.\nonumber
\end{eqnarray}
Its coproduct is given by
\begin{eqnarray}
\label{k22a}
\Delta \hat{D}_n & = & \hat{D}_n \otimes \Big(-ia \hat{D}_n+\sqrt{1-a^2\hat{D}_\mu\hat{D}_\mu}\Big)
 + \frac{ia
 \hat{D}_n+\sqrt{1-a^2\hat{D}_\mu\hat{D}_\mu}}{1-a^2\hdi_k\hdi_k}\otimes
 \hat{D}_n\nonumber \\
&&+ia\hdi_i\frac{ia \hat{D}_n+\sqrt{1-a^2\hat{D}_\mu\hat{D}_\mu}}{1-a^2\hdi_k\hdi_k}\otimes\hdi_i,\\
\Delta \hat{D}_j & = & \hat{D}_j\otimes \Big(-ia \hat{D}_n+\sqrt{1-a^2\hat{D}_\mu\hat{D}_\mu}\Big)
+  1\otimes \hat{D}_j.\nonumber
\end{eqnarray}
The Dirac derivative together with the generators of $SO_a(n)$ rotations
$M^{\mu\nu}$ forms a $\ka$-deformed Euclidean Hopf algebra which is
undeformed in the algebra sector, (\ref{k7a}) and (\ref{k20}). The deformation is purely
in the coalgebra sector, (\ref{k8}) and (\ref{k22a}). This
special basis of $SO_a(n)$  we will refer to in the
following as \emph{the} $SO_a(n)$. Recall that it is not unique (\ref{k20a}). Together with
the counit and the antipode of the Dirac derivative
\begin{eqnarray}
\label{k22b}
\epsilon(\hdi_n)&=& 0, \qquad S(\hdi_n)=-\hdi_n+ia \hdi_l\hdi_l\frac{ia
  \hat{D}_n+\sqrt{1-a^2\hat{D}_\mu\hat{D}_\mu}}{1-a^2\hdi_k\hdi_k},\nonumber\\
\epsilon(\hdi_j)&=& 0, \qquad S(\hdi_j)=-\hdi_j\frac{ia
  \hat{D}_n+\sqrt{1-a^2\hat{D}_\mu\hat{D}_\mu}}{1-a^2\hdi_k\hdi_k},
\end{eqnarray}
and the property $S^2(\hdi_\mu)=\hdi_\mu$, all operations of the
full Euclidean Hopf algebra $SO_a(n)$ have been
defined.

The square of the Dirac derivative  is not identical to the Laplace operator, but
$ \hat{D}_\mu\hat{D}_\mu=\hat{\Box}(1-\frac{a^2}{4}\hat{\Box})$.
However, having in mind the caveats below equations (\ref{k16}) and (\ref{k20a}) we
could rescale the Dirac derivative $\hdi_\mu'=
\frac{\hdi_\mu}{\sqrt{1-\frac{a^2}{4}\hat{\Box}}}$  such that
$\hdi_\mu'\hdi_\mu' =\hat{\Box}$. We do not follow this train of thought in
this article.

We quote  also the antipode $S(\hat{\Box})=\hat{\Box}$ of the
Laplace operator, its commutators with coordinates $\lb\hat{\Box}, \hat{x}^\mu\rb =  2\hat{D}_\mu$
and its Leibniz rule
\begin{eqnarray}
\label{k24}
\hat{\Box}\hspace{1mm} (\hat{f} \cdot \hat{g}) & = & (\hat{\Box} \hat{f})
\cdot (e^{-ia\hat{\partial}_n}\hat{g}) +
(e^{ia\hat{\partial}_n}\hat{f}) \cdot (\hat{\Box} \hat{g})+\nonumber\\
&  &+2(\hat{D}_ie^{ia\hat{\partial}_n}\hat{f})\cdot (\hat{D}_i \hat{g})+\frac{2}{a^2}((1-e^{ia\hat{\partial}_n})\hat{f})\cdot((1-e^{-ia\hat{\partial}_n})\hat{g}).
\end{eqnarray}
In (\ref{k24}) we have used the identities:
\begin{small}\begin{equation}
\label{k25}
e^{-ia\hat{\partial}_n} =-ia\hat{D}_n+
\sqrt{1-a^2\hat{D}_\mu\hat{D}_\mu} =1-ia\hat{D}_n-\frac{a^2}{2}\hat{\Box}, \hspace{5mm}
e^{ia\hat{\partial}_n}=\frac{ia\hat{D}_n+
  \sqrt{1-a^2\hat{D}_\mu\hat{D}_\mu}}{1-a^2\hat{D}_j\hat{D}_j}\hspace{1mm}.
\end{equation}\end{small}

There are also further invariants, such as in four-dimensional
$\ka$-Minkowski spacetime the Pauli-Lubanski vector, which has been discussed in \cite{lukrue2} and in
\cite{KLMSob2} in the bicrossproduct basis.
From
(\ref{k7a}) and (\ref{k20})  the generalisation of the
Pauli-Lubanski vector in $n=2m$ Euclidean dimensions can be deduced:
\begin{eqnarray}
\label{k26}
W^2_{i+1} &=& W_{\mu_1\dots\mu_{2i-1}}W_{\mu_1\dots\mu_{2i-1}},\hspace{8mm}W^2_{1}=\hdi_\mu\hdi_\mu ,\hspace{8mm}i=1,\dots\frac{n-2}{2} ,\nonumber\\
W_{\mu_1\dots\mu_{2i-1}} &=&
  \epsilon_{\mu_1\dots\mu_n}M^{\mu_{2i}\mu_{2i+1}}\dots M^{\mu_{n-2}\mu_{n-1}}\hat{D}_{\mu_n}.
\end{eqnarray}
All invariants should be identical to their undeformed
counterparts, exchanging ordinary
derivatives with the Dirac derivative. The reason is that the
algebra sector in
our particular basis of $SO_a(n)$ is undeformed.

%
%

\section{Star product and operator representations}

\subsection{Different $\star$-products}
In this section we
represent the associative algebra of functions of noncommuting coordinates as
an algebra of functions of commuting variables by means of  $\star$-products. This allows a
representation of the generators $\hdi_\mu$ and $M^{\mu\nu}$  of the
 Hopf algebra $SO_a(n)$ in terms of differential operators of
ordinary, commuting derivatives and coordinates.
The  representation by means of $\star$-products is particularly suitable, since it allows an expansion order by order in $a$.

The $\star$-product replaces the point-wise product of
commutative spacetime.
For a given noncommutative associative algebra of coordinates, there are
generically several different $\star$-products  fulfilling
\begin{equation}
\label{k33a}
\lb x^\mu \ds x^\nu \rb =x^\mu\star x^\nu -x^\nu \star x^\mu= i C^{\mu\nu}_\la x^\la, \quad
\textrm{if}\quad \lb \xx^\mu, \xx^\nu\rb =i C^{\mu\nu}_\la \xx^\la
\end{equation}
in the Lie algebra case for example.
Many interesting $\star$-products simply reproduce different ordering
prescriptions imposed on the abstract algebra of coordinates.  For
noncommutative coordinates the order in a monomial has to be specified, otherwise one would miscount the
elements of the basis of monomials. After multiplying two functions  of
noncommutative variables, the product has to be reordered according to
the chosen prescription. The 
$\star$-products of interest here perform this reordering for commuting
coordinates $x^\mu$.

In the case of
$\ka$-deformed space with only one noncommuting coordinate $\xx^n$, three
ordering prescriptions can be chosen in a generic way:
all factors of $\xx^n$ to the furthest left, all factors of $\xx^n$ to the
furthest right or complete symmetrisation of all coordinates.
We are especially interested in the symmetric $\star$-product because
of its hermiticity:
\begin{equation}
\label{k34}
\overline{f(x)\star g(x)}= \overline{g}(x) \star \overline{f}(x).
\end{equation}
Note that we
 set $f (x)\star g(x)\equiv (f\star g)(x)$, in other words we omit the
 multiplication map of the usual definition of the
 $\star$-product. This abuse of the usual notation allows to simplify the
 coproduct formulae in the following.

An important condition on a $\star$-product is that it is
associative. For Lie algebra models such as
$\ka$-deformed space with symmetric ordering, the
Baker-Campbell-Hausdorff (BCH) formula can be used to obtain an associative
symmetric $\star$-product. The BCH expansion involves the
structure constants of the Lie algebra (\ref{k4}):
\begin{eqnarray}
\label{k36}
f(x)\star g(x) &=& \exp\Big(\frac{i}{2}x^\la C^{\mu\nu}_\la \partial_\mu
\otimes  \partial_\nu  + \frac{1}{12}x^\la C^{\rho\si}_\la
C^{\mu\nu}_\rho (\partial_\si \otimes 1 - 1 \otimes \partial_\si) \partial_\mu
\otimes  \partial_\nu + \nonumber \\
&&\qquad + \frac{i}{24} x^\la
C^{\al\be}_\la C^{\rho\si}_\al C^{\mu\nu}_\rho \partial_\be
\partial_\mu\otimes \partial_\si\partial_\nu+\dots
\Big)\hspace{2mm} f (y) \otimes g (z)
|_{y,z\rightarrow x}\hspace{1mm}.
\end{eqnarray}
Generically  there is no closed symbolical
form, such as the Moyal-Weyl product, for $\star$-products of Lie algebra noncommutative spaces.
For the $\ka$-deformed space, however, there exists a closed formula, as we will show in the
following. This result has been found before \cite{KLM2}, \cite{ALZ}.
For our derivation we use the notion of equivalent
$\star$-products. Two $\star$-products $\star$ and $\star'$ are equivalent, if they can be
related by a differential operator $T$:
\begin{equation}
\label{k37}
T(f(x) \star g(x)) = T(f(x))\star' T(g(x)).
\end{equation}
For example, $\star$-products  which represent the same algebra with
a different ordering prescription are equivalent. We use this fact to
relate the symmetric $\star$-product to the  normal ordered
$\star$-products. These in turn can be derived via a Weyl quantisation
procedure (cp. \cite{madore}):
\begin{eqnarray}
\label{k38}
f(x)\star_L g(x) &=& \lim_{\substack{y\rightarrow x \\ z\rightarrow
    x}}\hspace{2mm}\exp\Big(x^j\frac{\p}{\p y^j}(e^{-ia\frac{\p}{\p z^n}}-1)\Big)\hspace{2mm}f(y)g(z),\nonumber\\
f(x)\star_R g(x) &=& \lim_{\substack{y\rightarrow x\\ z\rightarrow
    x}}\hspace{2mm}\exp\Big(x^j\frac{\p}{\p z^j}(e^{ia\frac{\p}{\p y^n}}-1)\Big)\hspace{2mm}f(y)g(z).
\end{eqnarray}
The $\star$-product $\star_L$ reproduces an ordering for
which all $\hat{x}^n$  stand on the \emph{L}eft hand side in
any monomial; $\star_R$ reproduces the opposite
ordering prescription. For our derivation we need several
formulae, which follow from (\ref{k38}) and from properties of the BCH formula (cp. \cite{Kathotia}):
\begin{eqnarray}
\label{k39}
x^j \star_L f(x) &=& x^j e^{-ia\p_n} f(x), \qquad\quad\hspace{6mm}  f(x) \star_L  x^j = x^j f(x),\nonumber\\
x^n \star_L  f(x) &=& x^n f(x),\qquad\qquad\quad\hspace{7mm}
 f(x)\star_L x^n   = \left(x^n
 -iax^k\p_k\right)f(x),\nonumber\\&&\nonumber\\
 x^j \star_R  f(x) &=& x^j  f(x), \qquad\qquad\qquad\quad  f(x) \star_R  x^j = x^j
 e^{ia\p_n} f(x),\nonumber\\
x^n \star_R  f(x) &=&\left(x^n
 +iax^k\p_k\right) f(x),\quad\hspace{2mm}
 f(x)\star_R x^n   =  x^n f(x).\nonumber\\&&\nonumber\\
 x^j \star  f(x) &=& x^j\frac{ia\p_n}{e^{ia\p_n}-1} f(x), \qquad\hspace{4mm} f(x)\star  x^j  = x^j\frac{-ia\p_n}{e^{-ia\p_n}-1}f(x),\\
\nonumber\\
x^n \star  f(x) &=& \left(x^n
 -\frac{x^k\p_k}{\p_n}(\frac{ia\p_n}{e^{ia\p_n}-1}-1)\right)f(x),\nonumber\\
 f(x)\star x^n  & =& \left(x^n
 -\frac{x^k\p_k}{\p_n}(\frac{-ia\p_n}{e^{-ia\p_n}-1}-1)\right)f(x).\nonumber
\end{eqnarray}

The operator $T$ up to first order in $a$ has to be of the form  $T=1+ia c
x^j\partial_j \partial_n +\dots$, with a real constant $c$ to be
determined. Note that $T$ depends also on the coordinates $x^j$. We
obtain $T(x^\mu)=x^\mu$ and for the left ordered $\star$-product $\star_L$
\begin{equation}
\label{k41}
T(x^j \star_L g(x)) = T(x^j)\star T(g(x))\hspace{2mm} \Rightarrow\hspace{2mm}  T(x^j e^{ia\p_n}
f(x)) = x^j \frac{ia\p_n}{e^{ia\p_n}-1} T(f(x)).
\end{equation}
Formula (\ref{k41}) can be written as  a differential equation with the
unique solution:
\begin{equation}
\label{k40}
T = \lim_{z\rightarrow x} \exp\Big(z^j\p_{x^j}(\frac{-ia\p_n}{e^{-ia\p_n}-1}-1)\Big),\hspace{5mm}
T^{-1} =\lim_{z\rightarrow x}\exp\Big(z^j\p_{x^j}(\frac{e^{-ia\p_n}-1}{-ia\p_n}-1)\Big),
\end{equation}
demanding that $T\cdot T^{-1}=1$.

 Similar equivalence operators $\tilde{T}$ can be
constructed relating $\star$ and $\star_R$, $\tilde{T}(f \star_R g) =
 \tilde{T}(f)\star \tilde{T}(g)$. The result is $\tilde{T} = \lim_{z\rightarrow x} \exp\Big(z^j\p_{x^j}(\frac{ia\p_n}{e^{ia\p_n}-1}-1)\Big)$.

With this solution for $T$ we  construct  the
symmetric $\star$-product:
{\small \begin{eqnarray}
\label{k42}
f(x)\star g(x)&=& \lim_{\substack{y\rightarrow x \\ z\rightarrow
    x}}T\Big(T^{-1}(f(z))\star_L T^{-1}(g(y))\Big)\nonumber\\
&=&\lim_{w\rightarrow
    x}\exp\Big(x^j\p_{w^j}(\frac{-ia\p_{w^n}}{e^{-ia\p_{w^n}}-1}-1)\Big)
    \lim_{\substack{z\rightarrow w \\ y\rightarrow
    w}}\hspace{2mm}\exp\Big(w^j
    \p_{z^j}(e^{-ia\p_{y^n}}-1)\Big)\cdot\\
&&\hspace{2mm}\lim_{\substack{u\rightarrow
    z \\ t\rightarrow y}}  \Big(\exp\Big(z^j\p_{u^j}(\frac{e^{-ia\p_{u^n}}-1}{-ia\p_{u^n}}-1)\Big)f(u)\Big)
    \Big(\exp\Big(y^j\p_{t^j}(\frac{e^{-ia\p_{t^n}}-1}{-ia\p_{t^n}}-1)\Big)g(t)\Big).\nonumber
\end{eqnarray}}
Contracting all limites, this result is written in a
compact way ($\partial_n=\pat_{y^n}+\pat_{z^n}$):
\begin{eqnarray}
\label{k43}
 f(x) \star  g(x) &=& \lim_{\substack{y\to x \\ z \to x}}\exp \Bigg( x^j \pat_{y^j}
   \left( \frac{\pat_n}{\pat_{y^n}} e^{-ia\pat_{z^n}} \frac{1-e^{-ia\pat_{y^n}}}{1-e^{-ia\pat_n}}
   - 1 \right)\nonumber\\
  & & \qquad\qquad + x^j \pat_{z^j}
   \left( \frac{\pat_n}{\pat_{z^n}} \frac{1-e^{-ia\pat_{z^n}}}{1-e^{-ia\pat_n}}- 1 \right)
   \Bigg) f(y)g(z).
\end{eqnarray}

%
%

\subsection{Representation of derivatives and generators of rotations}
\label{reps}
The symmetry algebra of
$\kappa$-deformed space, i.e. the generators of  rotations and the
Dirac derivatives, can be represented  as differential operators on
spaces of commuting variables with the symmetric $\star$-product as
multiplication:
\begin{eqnarray}
\label{k44}
\hat{\partial}_n \hat{f} \longrightarrow \partial_n^* f(x) & =&
  \partial_n  f(x),\nonumber\\
\hat{\partial}_j \hat{f} \longrightarrow \partial_j^* f(x) &=&
  \partial_j \Big(\frac{e^{ia\partial_n}-1}{ia\partial_n}\Big) f(x),
 \nonumber \\
N^l \hat{f}\longrightarrow N^{*l} f(x) & =& \Big( x^l\pat_n - x^n\pat_l
+ x^l\pat_\mu\pat_\mu \frac{e^{ia \pat_n}-1}{2\pat_n} \nonumber\\&&\qquad\quad- x^\mu\pat_\mu\pat_l
\frac{e^{ia\pat_n} -1- ia\pat_n}{ia\pat_n^2} \Big) f(x),\nonumber\\
M^{rs} \hat{f}\longrightarrow M^{*rs} f(x)& =& (x^s\pat_r-x^r\pat_s)f(x),\\
\hat{D}_n\hat{f} \longrightarrow D^*_n
  f(x)&=&\Big(\frac{1}{a}\sin(a\partial_n)-\frac{1}{ia\partial_n\partial_n}
  \partial_k\partial_k(\cos(a\partial_n)-1)\Big)f(x),\nonumber\\
\hat{D}_j\hat{f} \longrightarrow D^*_j
  f(x)&=& \partial_j \Big(\frac{e^{-ia\partial_n}-1}{-ia\partial_n}\Big) f(x),\nonumber\\
\hat{\Box}\hat{f} \longrightarrow \Box^* f(x)&=&\partial_\mu\partial_\mu\hspace{2mm}
\frac{2(1-\cos(a\partial_n))}{a^2\partial_n\partial_n}f(x).\nonumber
\end{eqnarray}
This representation can be derived in a perturbation expansion
on symmetrised monomials multiplied with the $\star$-product. However, it is
easier to derive it using the expressions $x^\mu \star
f(x)$ in (\ref{k39}). Rewriting these expressions symbolically as $x^{*\mu}f(x)$, relations such
as
\begin{equation}
\label{k46}
\lb \partial_j^*, x^{*n}\rb f(x)=\partial_j^* x^{*n}f(x)-
x^{*n}\partial_j^*f(x)\shouldid ia\partial_j^*f(x)
\end{equation}
have to be fulfilled for arbitrary $f(x)$. Note that all operators in  (\ref{k44}) coincide to zeroth order in $a$ with their
commutative counterparts.

Similar representations can be derived for the normal
ordered $\star$-products. These representations have to be different
from (\ref{k44}), since the  same abstract algebra of  $\hp_\mu$,
$\hdi_\mu$, $M^{rs}$  and $N^l$ is represented on different, but equivalent
$\star$-products. The different $\star$-representations can
be related with equivalence operators, via $\pat_j^{*_L}=T^{-1}\pat_j^{*}T$ etc.

For the left ordered $\star$-product ($\star_L$) we obtain:
\begin{eqnarray}
\label{k47}
\pat^{*_{L}}_n f(x) & = & \pat_n f(x),\nonumber\\
\pat_j^{*_{L}} f(x) & = & \pat_j e^{ia\pat_n}  f(x),\nonumber\\
N^{{*_{L}}\hspace{0.5mm}l}   f(x) & = & \Big( x^l\frac{1}{a} \sin (a\pat_n) - x^n\pat_l
e^{ia\pat_n}+\frac{ia}{2}x^l\p_k\p_k e^{ia\pat_n} \Big) f(x),\nonumber\\
M^{{*_{L}}\hspace{0.5mm}rs}   f(x) & = & \left(x^s\pat_r-x^r\pat_s \right)
f(x),\\
D^{*_{L}}_n   f(x) & = & \left( \frac{1}{a} \sin (a\pat_n) +
        \frac{ia}{2}\partial_k\partial_ke^{ia\pat_n}\right) f(x),\nonumber\\
D^{*_{L}}_j   f(x) & = & \partial_j f(x),\nonumber\\
\Box^{*_{L}}   f(x) &= &\left(
-\frac{2}{a^2} (\cos(a\pat_n) - 1)+\partial_k\partial_ke^{ia\pat_n} \right) f(x).\nonumber
\end{eqnarray}
The result for the right ordered $\star$-product ($\star_R$) is:
\begin{eqnarray}
\label{k48}
\pat^{*_{R}}_n f(x) & = & \pat_n f(x),\nonumber\\
\pat_j^{*_{R}} f(x) & = & \pat_j  f(x),\nonumber\\
N^{{*_{R}}\hspace{0.5mm}l}   f(x) & = & \Big( x^l\frac{1}{2ia} (e^{2ia\pat_n}-1) - x^n\pat_l
-iax^k\p_k\p_l+\frac{ia}{2}x^l\p_k\p_k  \Big) f(x),\nonumber\\
M^{{*_{R}}\hspace{0.5mm}rs}   f(x) & = & \left(x^s\pat_r-x^r\pat_s \right)
f(x),\\
D^{*_{R}}_n   f(x) & = & \left( \frac{1}{a} \sin (a\pat_n) +
        \frac{ia}{2}\partial_k\partial_ke^{-ia\pat_n}\right) f(x),\nonumber\\
D^{*_{R}}_j   f(x) & = & \partial_j e^{-ia\pat_n} f(x),\nonumber\\
\Box^{*_{R}}   f(x) &= &\left(
-\frac{2}{a^2} (\cos(a\pat_n) - 1)+\partial_k\partial_ke^{-ia\pat_n} \right) f(x).\nonumber
\end{eqnarray}

Summarising the ambiguities of our construction: we have
made the choice that the $SO_a(n)$ Hopf algebra is the one with
undeformed algebra sector.
Equation (\ref{k7}) followed uniquely from (\ref{k1}). In contrast, we have a
made an arbitrary choice following (\ref{k20a}), choosing the simplest
Dirac derivative out of a continuous range of solutions. The only other
choice is the  $\star$-product to represent the abstract
algebra. Hermiticity is a strong argument for favouring the symmetric
$\star$-product.

%
%
%
%
%
%
%

\section{Forms}

\subsection{Vector-like transforming one-forms}
A crucial ingredient of a geometric approach towards the
$\ka$-deformed space is the exterior differential, denoted by d. In order to find a representation of d, a
working definition of a one-form is needed.

The expected properties  of d are:
\begin{itemize}
\label{d1}
\item nilpotency: $\diff^2=0$;
\item application  of d to a coordinate gives a one-form $\lb \diff, \xx^\mu\rb=\hxi^\mu$;
\item invariance under  $SO_a(n)$: $\lb \hat{M}^{rs},
  \textrm{d}\rb =0$,  $\lb \hat{N}^{l},
  \textrm{d}\rb =0$;
\item undeformed Leibniz rule: $\diff (\hat{f}\cdot \hat{g})= (\diff
  \hat{f})\cdot \hat{g}+
  \hat{f}\cdot\diff \hat{g}$.
\end{itemize}
Demanding invariance of d under $SO_a(n)$, a natural ansatz is that the Dirac derivative $\hdi_\mu$ is the convenient
derivative dual to a set of vector-like transforming one-forms $\hxi^\mu$, d$=\hxi^\mu\hdi_\mu$:
\begin{equation}
\label{d10}
\lb \hat{M}^{rs},\hxi^\mu\rb =\de^{r\mu}\hxi^s-\de^{s\mu}\hxi^r, \qquad \lb \hat{N}^{l},\hxi^\mu\rb =\de^{n\mu}\hxi^l-\de^{l\mu}\hxi^n.
\end{equation}

The nilpotency of d$^2=0$ can be achieved requiring that one-forms  $\hxi^\mu$  commute with derivatives
and  anti-commute among themselves $\{\hxi^\mu, \hxi^\nu\}=0$.

Demanding that the commutator of d with a coordinate is a
one-form, $\lb \diff, \xx^\mu \rb =\hxi^\mu$ is a sufficient condition
for an undeformed Leibniz rule of d.

If we add the condition that the commutators $\lb \hxi^\mu, \xx^\nu \rb$ close in
the space of one-forms alone, there is no differential calculus consisting of $n$
one-forms fulfilling all these conditions simultaneously. Under this additional
condition,  a familiar result (e.g. \cite{Si1}) states  that the basis
of one-forms is $(n+1)$-dimensional. 

There have been hints towards this result in our discussion of the Dirac
operator (\ref{k22}).
Its commutator with the coordinates $\lb \hdi_\mu, \xx^\nu\rb$ is
a power series in the Dirac derivative alone, but it is linear
when adding the Laplace operator $\hat{\Box}$. The $(n+1)$-dimensional set of
derivatives ($\hdi_\mu$, $\hat{\Box}$) is dual to the $(n+1)$-dimensional set of one-forms
($\widehat{\textrm{d}x}^\mu$, $\widehat{\phi}$)
introduced in \cite{Si1}
\begin{eqnarray}
\label{d2}
\diff&=&\widehat{\textrm{d}x}^n
\hdi_n+\widehat{\textrm{d}x}^j\hdi_j-\frac{a^2}{2}\widehat{\phi}\hspace{1mm}\hat{\Box},\hspace{16mm}\lb  \diff, \xx^\mu \rb = \widehat{\textrm{d}x}^\mu,\nonumber
\nonumber\\
\lb \widehat{\textrm{d}x}^n, \xx^n \rb &=&a^2 \widehat{\phi},\qquad \hspace{3mm}\lb \widehat{\textrm{d}x}^j, \xx^n \rb =0,\hspace{6mm}\qquad\qquad\qquad\qquad \lb \widehat{\phi}, \xx^n \rb =-\widehat{\textrm{d}x}^n,\\
\lb \widehat{\textrm{d}x}^n, \xx^i \rb &=&ia \widehat{\textrm{d}x}^i,\qquad \lb \widehat{\textrm{d}x}^j, \xx^i \rb
=-ia\de^{ij}\widehat{\textrm{d}x}^n+a^2\de^{ij}\widehat{\phi},\qquad \lb \widehat{\phi}, \xx^i \rb
=-\widehat{\textrm{d}x}^i.\nonumber
\end{eqnarray}
It is a frequent observation in noncommutative geometry that the set
of one-forms on a particular space has one element more than in the
commutative setting. In our case this is acceptable at first sight
since for $a\rightarrow 0$, d
$\rightarrow \diff_{\textrm{\tiny class}}$. But several problems remain.
Only $n$ one-forms can be obtained by applying d to the
coordinates. A gauge theory with gauge potentials as
one-forms would result in an additional degree of freedom in the gauge
potentials. The cohomology of the differential calculus has an entirely
different structure than in the commutative case.

We will therefore follow a different strategy and demand as a central condition that there
are only $n$
one-forms on the noncommutative space. Of course we will not get
this condition for free, there will be a trade-off of the kind that
the one-forms $\hxi^\mu$ will have derivative-valued commutation
relations with the coordinates.   

To calculate the one-forms, we start from their commutators with
the coordinates:
\begin{eqnarray}
\label{d3}
\hxi^\nu\shouldid\lb \diff, \xx^\nu \rb &=& \lb \hxi^\mu\hdi_\mu, \xx^\nu\rb=
\nonumber\\
&=& \lb \hxi^\mu, \xx^\nu\rb \hdi_\mu + ia\hxi^n\hdi_\nu +\hxi^\nu\big(-ia\hdi_n +\sqrt{1-a^2\hdi_\mu\hdi_\mu}\big)\big).
\end{eqnarray}
From (\ref{d3}) follows that the commutator $\lb
\hxi^\mu,\xx^
\nu\rb$ will involve  derivatives. To calculate it
we have made a general ansatz with \emph{derivative-valued}
commutators $\lb
\hxi^\mu, \xx^\nu\rb$ involving all
terms compatible with the index structure. The result is derived
requiring that the commutators $\lb
\hxi^\mu, \xx^\nu\rb$ are compatible with (\ref{k1}). Invariance under
$SO_a(n)$ rotations does not add
further constraints and we find the unique solution:
\begin{equation}
\label{d4}
 \lb \hxi^\mu, \xx^\nu\rb= ia (\de^{\mu n}\hxi^\nu
 -\de^{\mu\nu}\hxi^n)+(\hxi^\mu\hdi_\nu+\hxi^\nu\hdi_\mu
 -\de^{\mu\nu}\hxi^\rho\hdi_\rho)\frac{1-\sqrt{1-a^2\hdi_\si
    \hdi_\si}}{\hdi_\la \hdi_\la}.
\end{equation}
As an aside note that $\frac{1-\sqrt{1-a^2\hdi_\si
    \hdi_\si}}{\hdi_\la \hdi_\la}
= \frac{a^2}{2}\frac{1}{1-\frac{a^2}{4}\hat{\Box}}$.

The price we have to pay for having only $n$ one-forms is that the commutator (\ref{d4}) is highly
    nonlinear in the Dirac derivative.

It is possible to generalise one of the conditions for the
differential calculus, the undeformed Leibniz rule $\lb \diff, \xx^\mu \rb =\hxi^\mu$.
We define commutation relations between a
second set of
one-forms $\widetilde{\xi}^\mu$ and coordinates $\xx^\nu$,
consistent with (\ref{k1}) and (\ref{k7a}). For these one-forms
$\widetilde{\xi}^\mu$ the application of d to a
coordinate does not return the one-form, but a derivative-valued expression
\begin{equation}
\label{d7}
\lb \widetilde{\xi}^n\hdi_n+\widetilde{\xi}^j\hdi_j, \xx^\nu\rb=\lb \diff, \xx^\nu\rb= (\diff\xx)^\nu=\widetilde{\xi}^\nu\cdot f(\hdi_n,\hdi_j\hdi_j),
\end{equation}
with a suitable function of the Dirac derivative
$f(\hdi_n,\hdi_j\hdi_j)$.

The most general solution for (\ref{d7}) is
\begin{eqnarray}
\label{d8}
\lb \diff,\xx^\nu\rb &=& \widetilde{\xi}^\nu+\hspace{1mm}c'\hspace{1mm}\widetilde{\xi}^\nu\big(\frac{1}{\sqrt{1-a^2\hdi_\si\hdi_\si}}-1\big)\nonumber\\
 \lb \widetilde{\xi}^\mu, \xx^\nu\rb&=& ia (\de^{\mu n}\widetilde{\xi}^\nu
 -\de^{\mu\nu}\widetilde{\xi}^n)+(1-c')\big(\widetilde{\xi}^\mu\hdi_\nu+\widetilde{\xi}^\nu\hdi_\mu
 -\de^{\mu\nu}\widetilde{\xi}^\rho\hdi_\rho\big)\frac{1-\sqrt{1-a^2\hdi_\si
    \hdi_\si}}{\hdi_\la \hdi_\la}\nonumber\\
    &&+\hspace{1mm}c'\hspace{1mm}\widetilde{\xi}^\nu\hdi_\mu \frac{a^2}{\sqrt{1-a^2\hdi_\la\hdi_\la}},
\end{eqnarray}
for an arbitrary constant $c'$.  The solution (\ref{d4}) corresponds to
$c'=0$ and we will use it exclusively in the following.

With (\ref{d4}) it is very difficult to calculate the action of d on a general
$x$-dependent one-form $\al_\mu(\xx)\hxi^\mu$:
\begin{equation}
\label{d11}
\diff \al=\diff (\al_\mu(\xx)\hxi^\mu)=\hxi^\nu (\hdi_\nu
\al_\mu(\xx)) \hxi^\mu \neq (\hdi_\nu
\al_\mu(\xx))\hxi^\nu \hxi^\mu.
\end{equation}
However, a general one-form may be defined in such a way that
$\hxi^\mu$ stands to the left of the coefficient function:
\begin{equation}
\label{d11a}
\diff \al=\diff (\hxi^\mu\al_\mu(\xx))=\hxi^\nu \hxi^\mu (\hdi_\nu
\al_\mu(\xx)).
\end{equation}
Still it is interesting to see whether there are one-forms which allow an action of d as in
(\ref{d11}), independent of the order. This motivates the introduction of a second basis of one-forms, which we
call frame.
%
%

\subsection{Representation of $\hxi^\mu$}
In analogy with the approach in section \ref{reps}
we now derive a representation of the one-forms
$\hxi^\mu \rightarrow \xi^{*\mu}$ in the $\star$-product setting.

The starting point  is the commutator (\ref{d4}), which involves
 a power series expansion in the derivatives. This implies that
the $\xi^{*\mu}$ can be written as functions of the commutative
one-forms $\diff x^\mu$ and the commutative derivatives
$\partial_\nu$.  The one-forms $\xi^{*\mu}$ should be at most linear in  $\diff x^\mu$.

We make the most general ansatz compatible with the index structure:
\begin{eqnarray}
\label{d30}
\xi^{*n}&=& \diff x^n e_1(\pat_i \pat_i, \pat_n) + \diff x^k \pat_k
e_2 (\pat_i \pat_i, \pat_n),\nonumber\\
\xi^{*j}&=& \diff x^j f_1(\pat_i \pat_i, \pat_n) + \diff x^n \pat_j
f_2 (\pat_i \pat_i, \pat_n)+ \diff x^k\pat_k \pat_j
f_3 (\pat_i \pat_i, \pat_n).
\end{eqnarray}

We have stated in (\ref{k46}) how to calculate the representation of a
derivative operator from its commutator with a coordinate.
We collect the terms
proportional to different one-forms $\diff x^\mu$ and different combinations
of derivatives. We obtain an overdetermined system of equations which can be solved
consistently. With the
abbreviation
\begin{equation}
\label{d34}
\gamma=\frac{1}{1+\frac{\pat_\mu\pat_\mu}{2\pat_n^2}(\cos(a\pat_n)-1)},
\end{equation}
we obtain the result
\begin{eqnarray}
\label{d35}
f_1&=&\gamma,\hspace{44mm}e_1=\Big(1+\cos(a\pat_n)-\frac{\pat_k\pat_k}{\pat_n^2}(\cos(a\pat_n)-1)\Big)\gamma^2,\nonumber\\
f_2&=&-\frac{2i}{\pat_n}\sin(a\pat_n)\gamma^2,\hspace{20mm}e_2=\frac{2i}{\pat_n}\sin(a\pat_n)\gamma^2,\\
f_3&=&-\frac{1}{\pat_n^2}(\cos(a\pat_n)-1)\gamma^2,\nonumber
\end{eqnarray}
or
{\small \begin{eqnarray}
\label{d36}
\xi^{*n}&=& \Big( \diff x^n
(1+\cos(a\pat_n)-\frac{\pat_k\pat_k}{\pat_n^2}(\cos(a\pat_n)-1)+\diff x^k
\frac{2i\pat_k}{\pat_n}
\sin(a\pat_n)\Big)\gamma^2,\\
\xi^{*j}&=& \Big( \diff x^j (1+\frac{\pat_\mu\pat_\mu}{2\pat_n^2}(\cos(a\pat_n)-1))-\diff x^n
\frac{2i\pat_j}{\pat_n}\sin(a\pat_n)-\diff x^k\frac{2\pat_k\pat_j}{\pat^2_n}(\cos(a\pat_n)-1)\Big)\gamma^2.\nonumber
\end{eqnarray}}

The more general differential calculus (\ref{d8}) has a particularly
simple solution for $c'=1$. The one-forms
$\widetilde{\xi}^\mu$ for $c'=1$ have the
following $\star$-representation $\tilde{\xi}^{*\mu}$:
{\small \begin{eqnarray}
\label{d37}
\widetilde{\xi}^{*n}&=& \Big(\diff x^n +\diff x^l
\frac{\pat_l}{\pat_n}(1-e^{-ia\pat_n})\Big)\frac{1}{1+\frac{\pat_\mu\pat_\mu}{2\pat_n^2}(\cos(a\pat_n)-1)},\\
\widetilde{\xi}^{*j}&=& \Big(-\diff x^l\frac{\pat_l\pat_j}{\pat_n\pat_n}(\cos(a\pat_n)-1) +\diff x^n
\frac{\pat_j}{\pat_n}(1-e^{-ia\pat_n})\Big)\frac{1}{1+\frac{\pat_\mu\pat_\mu}{2\pat_n^2}(\cos(a\pat_n)-1)}.\nonumber
\end{eqnarray}}
It is interesting to note that for this specific set of
differentials $\widetilde{\xi}^{*\mu}$ we obtain a representation on the space
of functions multiplied with the $\star$-product, where
$\widetilde{\xi}^{*j}$ is not proportional to  $\diff x^j$.

%
%

\subsection{Frame one-forms}
We have defined
one-forms $\hat{\xi}^\mu$ with vector-like transformation behaviour under
$SO_a(n)$. 
Alternatively we can define one-forms starting from the condition that
they commute with
coordinates $\lb \hat{\omega}^\mu, \xx^\nu\rb=0$ and therefore with all
functions. We make the ansatz
\begin{eqnarray}
\label{d50}
\hat{\xi}^n&=&\hat{\omega}^n g_1(\hat{D}_n,
\hat{D}_l\hat{D}_l)+\hat{\omega}^j\hat{D}_j g_2(\hat{D}_n,
\hat{D}_l\hat{D}_l),\\
\hat{\xi}^i&=&\hat{\omega}^n\hat{D}_i h_1(\hat{D}_n,
\hat{D}_l\hat{D}_l)+\hat{\omega}^i h_2(\hat{D}_n,
\hat{D}_l\hat{D}_l)+\hat{\omega}^j\hat{D}_j\hat{D}_i h_3(\hat{D}_n,
\hat{D}_l\hat{D}_l),\nonumber
\end{eqnarray}
with  functions of the Dirac derivative with appropriate index structure
and  expand  (\ref{d4}).
Since we assume that $\hom^\mu$ commute with the coordinates, we can reduce
the result
to commutators of the functions of derivatives $g_a$ and $h_a$ with
the coordinates. With the abbreviations
\begin{equation}\label{d57y}
\zeta_1=\frac{1}{\sqrt{1-a^2\hdi_\mu\hdi_\mu}},\hspace{5mm}
\quad
\zeta_2=\frac{1-\sqrt{1-a^2\hdi_\mu\hdi_\mu}}{\hdi_\si\hdi_\si}, \hspace{5mm}
\zeta_3=-ia\hdi_n+\sqrt{1-a^2\hdi_\mu\hdi_\mu},
\end{equation}
and the identities
\begin{equation}\label{d57z}
\frac{\partial}{\partial  \hdi_n}\zeta_2=\hdi_n\zeta_1\zeta_2^2,\hspace{8mm}
\frac{\partial}{\partial  \hdi_j\hdi_j}\zeta_2=\frac{1}{2}\zeta_1\zeta_2^2,\hspace{8mm}
\frac{\partial}{\partial \hdi_n}\zeta_3=-ia\zeta_1\zeta_3,
\end{equation}
equation (\ref{d4}) can be rewritten as a system of differential
equations for $g_1$, $g_2$, $h_1$, $h_2$ and $h_3$.
Its solution is
\begin{eqnarray}
\label{d58}
g_1&=& (1+\hdi_j\hdi_j\zeta_2\zeta_3^{-1})\frac{a^2}{2}\zeta_2,\hspace{15mm}
h_1=(-ia-\hdi_n\zeta_2)\frac{a^2}{2}\zeta_2\zeta_3^{-1},\nonumber\\
g_2&=&(ia+\hdi_n\zeta_2)\frac{a^2}{2}\zeta_2\zeta_3^{-1},\hspace{18mm}
h_2=\frac{a^2}{2}\zeta_2,\\
&&\hspace{53.5mm}h_3=\frac{a^2}{2}\zeta^2_2\zeta_3^{-1}.\nonumber
\end{eqnarray}
Writing the differential $\td$ in terms of the frame
one-forms $\hom^\mu$ we obtain
{\small \begin{eqnarray}
\label{d59}
\td&=&\hxi^\mu\hdi_\mu=
\Big(\hom^n\hdi_n-ia\hom^n\hdi_l\hdi_l\zeta_3^{-1}+\hom^j\hdi_j\zeta_3^{-1}\Big)\frac{a^2}{2}\zeta_2\nonumber\\
&=& \Big(\hom^n\hdi_n+\frac{\hom^j\hdi_j-ia\hom^n\hdi_l\hdi_l}{-ia\hdi_n+\sqrt{1-a^2\hdi_\mu\hdi_\mu}}\Big)\frac{a^2}{2}\frac{1-\sqrt{1-a^2\hat{D}_\si\hat{D}_\si}}{\hat{D}_\la\hat{D}_\la}\hspace{2mm}.
\end{eqnarray}}
We can simplify this result using the Laplace operator
$\hat{\Box}$ and the derivatives $\hp_\mu$
\begin{equation}
\label{d60}
\td=\Big(\hom^n(\frac{1}{a}\sin(a\hp_n)-\frac{ia}{2}\hp_l\hp_l
e^{-ia\hp_n})+\hom^j \hp_j\Big)\frac{1}{1-\frac{a^2}{4}\hat{\Box}}\hspace{2mm}.
\end{equation}

To determine the transformation behaviour of
$\hom^\mu$ under $SO_a(n)$-rotations,
we first consider the derivative operators dual to  $\hom^\mu$. The factor $\frac{1}{1-\frac{a^2}{4}\hat{\Box}}$ is an
invariant under $SO_a(n)$-rotations by itself. We define
\begin{equation}
\label{d61}
\widetilde{\p}_j=\hp_j, \qquad \widetilde{\p}_n=\frac{1}{a}\sin(a\hp_n)
-\frac{ia}{2}\hp_j\hp_j e^{-ia\hp_n}.
\end{equation}
By means of (\ref{k12}) we can calculate
\begin{eqnarray}
\label{d62}
\lb M^{rs},\widetilde{\p}_j\rb&=&\de^{r}_{j}\widetilde{\p}_s-\de^{s}_{j}\widetilde{\p}_r,\nonumber\\\lb
M^{rs},\widetilde{\p}_n\rb&=&0,\nonumber\\
\lb
N^{l},\widetilde{\p}_j\rb&=&-\de^{l}_{j}\widetilde{\p}_n\sqrt{1-a^2\widetilde{\p}_\mu\widetilde{\p}_\mu}-ia\de^{l}_{j}\widetilde{\p}_\mu\widetilde{\p}_\mu+ia\widetilde{\p}_j\widetilde{\p}_l,\\
\lb
N^{l},\widetilde{\p}_n\rb&=&\widetilde{\p}_l(ia \widetilde{\p}_n+\sqrt{1-a^2\widetilde{\p}_\mu\widetilde{\p}_\mu}).\nonumber
\end{eqnarray}
The derivatives $\widetilde{\p}_\mu$ are a
module of $SO_a(n)$ and $\lb N^l,
\widetilde{\p}_\mu\widetilde{\p}_\mu \rb = 0$. Comparing with
(\ref{k25}) we find:
\begin{equation}
\label{d63}
e^{-ia\hp_n}
=\frac{-ia\widetilde{\p}_n+\sqrt{1-a^2\widetilde{\p}_\mu\widetilde{\p}_\mu}}{1-a^2\widetilde{\p}_k\widetilde{\p}_k},\hspace{10mm}
e^{ia\hp_n}
=ia\widetilde{\p}_n+\sqrt{1-a^2\widetilde{\p}_\mu\widetilde{\p}_\mu},
\end{equation}
and the coproducts are
\begin{eqnarray}
\label{d64}
\widetilde{\p}_j(\hat{f}\cdot\hat{g})&=&\widetilde{\p}_j(\hat{f})\cdot\hat{g}+\big(e^{ia\hp_n}\hat{f}\big)\cdot (\widetilde{\p}_j\hat{g}),\nonumber\\
\widetilde{\p}_n(\hat{f}\cdot\hat{g})&=&\widetilde{\p}_n(\hat{f})\cdot(e^{-ia\hp_n}\hat{g})+\big(e^{ia\hp_n}\hat{f}\big)\cdot (\widetilde{\p}_n\hat{g})-ia(\widetilde{\p}_k\hat{f})\cdot(e^{-ia\hp_n}\widetilde{\p}_k\hat{g}).
\end{eqnarray}
For compactness, we have used (\ref{d63}) to write (\ref{d64}).
We find
$\widetilde{\p}_\mu\widetilde{\p}_\mu=\hdi_\mu\hdi_\mu$. Therefore the
Laplace operator $\hat{\Box}$ can be written
in terms of $\widetilde{\p}_\mu$ as $\hat{\Box}=\frac{2}{a^2}(1-\sqrt{1-a^2\widetilde{\p}_\mu\widetilde{\p}_\mu})
$. Note that $\widetilde{\p}_\mu=S(\hdi_\mu)$, the antipode of the
Dirac derivative.
Thus, we can  write (\ref{d60}) purely in terms of $\hom^\mu$
and $\widetilde{\p}_\mu$:
\begin{equation}
\label{d66}
\td=\Big(\hom^n\widetilde{\p}_n+\hom^j \widetilde{\p}_j\Big)\frac{2}{1+\sqrt{1-a^2\widetilde{\p}_\mu\widetilde{\p}_\mu}}.
\end{equation}
Comparing with (\ref{d11}), we can now evaluate the action of the differential
on an arbitrarily ordered one-form
\begin{equation}
\label{d66a}
\diff \al=\diff (\al_\mu(\xx)\hom^\mu)= \Bigg(\frac{2\widetilde{\p}_\nu}{1+\sqrt{1-a^2\widetilde{\p}_\la\widetilde{\p}_\la}}
\al_\mu(\xx)\Bigg) \hom^\nu \hom^\mu.
\end{equation}

From (\ref{d62}) we can calculate the transformation behaviour of
$\hom^\mu$ from the requirement that d is an invariant:
\begin{eqnarray}
\label{d67}
\lb M^{rs}, \hom^n\rb&=&0,\hspace{25mm}\lb N^{l}, \hom^n\rb=
\hom^l\sqrt{1-a^2\widetilde{\p}_\mu\widetilde{\p}_\mu} +ia
(\hom^l\widetilde{\p}_n-\hom^n\widetilde{\p}_l),\\
\lb M^{rs}, \hom^j\rb&=& \de^{rj}\hom^s-\de^{sj}\hom^{r},\hspace{5mm}
\lb N^{l}, \hom^j\rb=-\de^{lj}
\hom^n\sqrt{1-a^2\widetilde{\p}_\mu\widetilde{\p}_\mu} +ia
(\hom^l\widetilde{\p}_j-\hom^j\widetilde{\p}_l).\nonumber
\end{eqnarray}
The frame one-forms form a module of $SO_a(n)$ rotations.

The commutation relations between derivatives
$\widetilde{\p}_\mu$
and  coordinates are
\begin{eqnarray}
\label{d68}
\lb \widetilde{\p}_j, \xx^n \rb &=&ia\widetilde{\p}_j,\hspace{15mm}\lb \widetilde{\p}_n, \xx^n \rb =\frac{-ia^3\widetilde{\p}_s\widetilde{\p}_s\widetilde{\p}_n+\sqrt{1-a^2\widetilde{\p}_\mu\widetilde{\p}_\mu}}{1-a^2\widetilde{\p}_k\widetilde{\p}_k} ,\nonumber\\
\lb \widetilde{\p}_j, \xx^i \rb &=&\de^i_j ,\hspace{19mm}
\lb \widetilde{\p}_n, \xx^i \rb =-ia\widetilde{\p}_i\frac{-ia\widetilde{\p}_n+\sqrt{1-a^2\widetilde{\p}_\mu\widetilde{\p}_\mu}}{1-a^2\widetilde{\p}_k\widetilde{\p}_k} .
\end{eqnarray}

Taking into account 
$\frac{2}{1+\sqrt{1-a^2\widetilde{\p}_\mu\widetilde{\p}_\mu}}$
in the commutator with the coordinates, we define
$\eth_\nu=\frac{2\widetilde{\p}_\nu}{1+\sqrt{1-a^2\widetilde{\p}_\mu\widetilde{\p}_\mu}}$
as the derivatives dual to $\hom^\mu$.
Using
$\frac{2}{1+\sqrt{1-a^2\widetilde{\p}_\mu\widetilde{\p}_\mu}}=1+\frac{a^2}{4}\eth_\mu\eth_\mu$,
we obtain
{\small\begin{eqnarray}
\label{d69}
\lb \eth_j,
\xx^n \rb &=&\frac{ia}{2} \eth_j (1+\frac{a^2}{4}\eth_\mu\eth_\mu)
\Big(1+\frac{(1-ia\eth_n-\frac{a^2}{4}\eth_\nu\eth_\nu)
(1+\frac{a^2}{4}\eth_\la\eth_\la)}
{1+\frac{a^2}{4}\eth_\rho\eth_\rho-a^2 \eth_k\eth_k}\Big),\nonumber\\
\lb \eth_j,\xx^i \rb &=&
\de^i_j (1+\frac{a^2}{4}\eth_\mu\eth_\mu)+\frac{a^2}{2}\eth_i\eth_j\cdot
\frac{1-ia\eth_n-\frac{a^2}{4}\eth_\nu\eth_\nu}
{1+\frac{a^2}{4}\eth_\rho\eth_\rho-a^2 \eth_k\eth_k},\nn\\
\lb \eth_n,\xx^i \rb &=&
-\frac{ia}{2} \eth_i \Big(1+\frac{1-\frac{a^2}{4}\eth_\nu\eth_\nu}
{1+\frac{a^2}{4}\eth_\la\eth_\la}+\frac{ia}{2}\eth_n
\frac{1}{1+\frac{a^2}{4}\eth_\rho\eth_\rho}\Big),\\
\lb \eth_n,\xx^n \rb &=&(1+\frac{a^2}{4}\eth_\mu\eth_\mu)\cdot
\Big(1+\frac{-\frac{ia^3}{8}\eth_s\eth_s\eth_n+(1-\frac{a^2}{4}\eth_\mu\eth_\mu)
(1+\frac{a^2}{4}\eth_\nu\eth_\nu)^2}
 {(1+\frac{a^2}{4}\eth_\rho\eth_\rho)^2(1+(\frac{a^2}{4}\eth_\sigma\eth_\sigma)^2
 +\frac{a^2}{2}(\eth_n\eth_n-\eth_k\eth_k)}\Big).\nn
\end{eqnarray}}
These complicated commutators are the price we have to pay for the
fact that frame one-forms commute with all functions.

%
%

\subsection{Volume form}
The result (\ref{d4}) allows to calculate the
commutation relations of higher-order forms
$\hxi^{\mu_1}\dots\hxi^{\mu_j}$ with the coordinates. We can therefore
calculate  two-forms, three-forms etc. up to $n$-forms, the full Hodge
differential calculus.

  Since we know that the $\hxi^\mu$ anti-commute among
themselves, the dimension of the set of $j$-forms is
$\binom{n}{j}$. From the vector-like transformation  behaviour of
$\hxi^\mu$ (\ref{d10}) follows that $j$-forms
transform as $j$-tensors. Specifically, there is only one $n$-form $\hxi^{1}\hxi^{2} \dots
 \hxi^{n}$,
which should be a noncommutative analog of the volume form.
It has particularly simple properties, from (\ref{d4})  we can calculate
the commutator
\begin{equation}
\label{d40}
\lb \hxi^{1}\hxi^{2} \dots
 \hxi^{n}, \xx^\mu \rb = n\hxi^{1}\hxi^{2} \dots
 \hxi^{n}\hdi_\mu \frac{1-\sqrt{1-a^2\hdi_\si \hdi_\si}}{\hdi_\la \hdi_\la}.
\end{equation}
The volume form $\hxi^{1}\hxi^{2} \dots
 \hxi^{n}$ is invariant under $SO_a(n)$: $\lb M^{\mu\nu},\hxi^{1}\hxi^{2} \dots
 \hxi^{n}\rb=0$.

In contrast to this result, the $n$-form built from $n$ different
frame one-forms $\hom^1\hom^2\dots \hom^n$ is not invariant under
$SO_a(n)$:  $\lb
N^{l},\hom^1\hom^2\dots\hom^n\rb=-ia(n-1)\hom^1\hom^2\dots\hom^n\hp_l$.

The $\star$-representation of the volume form  is:
\begin{eqnarray}
\label{d41}
\big(\xi^{1}\xi^2\dots\xi^n\big)^*= \frac{\diff x^1  \diff
  x^2 \dots  \diff x^n}{\Big(1+\frac{\pat_\mu\pat_\mu}{2\pat^2_n}(\cos(a\pat_n)-1)\Big)^n},
\end{eqnarray}
while the representation of the frame $n$-form is simply $\big(\omega^1\omega^2\dots\omega^n\big)^{*}=\diff x^1  \diff
  x^2 \dots  \diff x^n$.

%
%
%
%
%
%
%

\section{Vector fields}
\subsection{Linearly transforming vector fields and conjugation}
The aim of our work is to define physical field theories. Vector fields that have the same transformation properties as
derivatives under $SO_a(n)$ are a
necessary ingredient for the definition of gauge
theories.

The central assumption is that the transformation
behaviour of vector fields is such that the vector fields appear linearly on the right
hand side of the commutation relations.

Vector fields analogous to the vector-like
transforming Dirac derivative
are obviously
\begin{eqnarray}
\label{v1}
\lb M^{rs},\hat{V}_n\rb &=& 0,\qquad\hspace{2.5mm}
\lb M^{rs},\hat{V}_i\rb =\delta _i^r\hat{V}^s-\delta _i^s\hat{V}^r, \nonumber \\
\lb \hat{N}^l,\hat{V}_n\rb &=& \hat{V}^l, \quad\quad
\lb \hat{N}^l,\hat{V}_i\rb = -\delta _i^l\hat{V}_n,
\end{eqnarray}
these vector fields $\hat{V}_\mu$ are a module of  $SO_a(n)$ rotations.

It is more
difficult to find vector fields with transformation properties
analogous to the other derivatives that we have defined throughout
this paper. Although we have argued that the  derivatives $\hp _\mu$
are in a sense irrelevant for the geometric construction of
$\ka$-deformed space, they have an important role to play for making
contact with the commutative regime. Since $\hp_n=\pat_n$ on all three $\star$-products, these derivatives
$\hp_\mu$ provide information on the connection between the abstract
algebra and $\star$-product
representation. Therefore we now investigate vector fields
$\hat{A}_\mu$  analogous to $\hp_\mu$. By this we mean that we
construct the tramsformation law of  $\hat{A}_\mu$ in such a way that
it conicides with  (\ref{k12}), when $\hat{A}_\mu$
is re-substituted with $\hat{\partial}_\mu$. At the same time $\hat{A}_\mu$
must be a module of
$SO_a(n)$ rotations.  We make the choice that  derivatives are always to
the left  of the vector field $\hat{A}_\mu$ in nonlinear expressions
such as the vector field analog of (\ref{k12}). We stress that
$\hat{A}_\mu$ is treated as an element of an abstract algebra in this
approach. Therefore derivatives  are not evaluated on
$\hat{A}_\mu$ in terms of the coproduct.

The problem can be solved in a
power series expansion in $a$. This results in a recursion
formula with the solution\footnote{This solution is not unique. If the
  symmetrisation in the third term of $\lb N^l, \hat{A}_i\rb$ is not
  performed, the last term of $\lb N^l, \hat{A}_i\rb$ vanishes.}:
\begin{eqnarray}
\label{v2}
\lb M^{rs}, \hat{A}_i\rb &=& \delta ^r_i \hat{A}_s -  \delta ^s_i
\hat{A}_r, \qquad\quad\quad
\lb M^{rs}, \hat{A}_n\rb = 0, \nonumber\\
\lb N^l, \hat{A}_i\rb &=& \delta ^l_i\frac{1-e^{2ia\hp _n}}{2ia\hp _n}\hat{A}_n -\frac{ia}{2}\delta^ l_i \hp _j\hat{A}_j+ \frac{ia}{2}\big( \hp _l\hat{A}_i +\hp _i\hat{A}_l\big)
\nonumber\\
&&-\delta ^l_i\frac{a}{2\hp _n}\tan \big(\frac{a\hp _n}{2}\big)\big(
\hp _n\hp _j\hat{A}_j-\hp _j\hp _j\hat{A}_n\big) \\
&&+\Big( \frac{1}{\hp _n^2}
-\frac{a}{2\hp _n}\cot \big(\frac{a\hp _n}{2}\big)\Big)\big( \hp _n\hp _i\hat{A}_l+\hp _n\hp _l\hat{A}_i
-2\hp _l\hp _i\hat{A}_n \big) ,\nonumber\\
\lb N^l, \hat{A}_n\rb &=& \hat{A}_l. \nonumber
\end{eqnarray}

The square of the vector field corresponding to the Dirac derivative is
an invariant $\lb M^{\mu\nu}, \hat{V}_\la\hat{V}_\la\rb=0$. To form an
invariant  out of the vector field $\hat{A}_\mu$, we have to
define a vector
field $\Breve{A}_\mu$  with
transformation laws in which the derivatives
are to the right of the vector
field $\Breve{A}_\mu$. We demand
\begin{equation}
\label{v3}
\lb M^{rs},\Breve{A}_\la\hat{A}_\la\rb =0,\quad \textrm{ and  }\quad
\lb N^{l},\Breve{A}_\la\hat{A}_\la\rb =0.
\end{equation}
From (\ref{v2})
we can construct the transformation laws for $\Breve{A}_\mu$ such that
(\ref{v3}) is fulfilled:
\begin{eqnarray}
\label{v4}
\lb M^{rs}, \Breve{A}_i\rb &=& \delta ^r_i \Breve{A}_s -
\delta ^s_i \Breve{A}_r, \qquad\quad\quad
\lb M^{rs}, \Breve{A}_n\rb = 0, \nonumber\\
\lb N^l, \Breve{A}_i\rb &=& -\delta ^l_i\Breve{A}_n
+{ia\over 2}\Breve{A}_l\hp _i
-{ia\over 2}\Breve{A}_i\hp _l
-{ia\over 2}\delta^l_i\Breve{A}_j\hp _j \nonumber\\
&&+ {a\over 2}\Breve{A}_l\tan\big(\frac{a\hp _n}{2}\big)\hp _i
-(\delta^l_i\Breve{A}_j\hp _j+\Breve{A}_i\hp _l)
\Big( {1\over \hp _n}-{a\over 2}\cot\big( \frac{a\hp _n}{2}\big)\Big)  ,\\
\lb N^l, \Breve{A}_n\rb &=& \Breve{A}_l{e^{2ia\hp _n}-1\over 2ia\hp _n}
-\Breve{A}_l{a\over 2\hp _n}\tan\big({a\hp _n\over 2}\big)\hp _j\hp _j + 2\Breve{A}_j\Big( {1\over \hp _n^2}
-{a\over 2\hp _n}\cot\big( {a\hp _n\over 2}\big)\Big)\hp _l \hp _j.\nonumber
\end{eqnarray}
With this transformation law the vector fields $\Breve{A}_\mu$ are a module of
$SO_a(n)$ rotations.

All relations considered up to now are invariant
under the conjugation
\begin{eqnarray}
\label{v5}
(\hat{x}^\mu)^\dagger=\hat{x}^\mu, && \quad\quad (\hp _\mu)^\dagger=-\hp _\mu ,\nonumber\\
(M^{rs})^\dagger=-M^{rs}, && \quad\quad (N^l)^\dagger=-N^l.
\end{eqnarray}
Comparing (\ref{v2}) and (\ref{v4}),  we see that
$\Breve{A}_\mu$ transforms with the derivatives
on the right hand side, but $\Breve{A}_\mu^\dagger \ne \hat{A}_\mu$, they
transform in different ways.
The transformation for $\hat{A}_\mu^\dagger$ is simply (\ref{v2}),
with all $\hat{A}_\mu$ standing to the furthest right in any
expression replaced
by $\hat{A}_\mu^\dagger$ standing to the furthest left.

The dual of $\hat{A}^\dagger_\mu$ is $\Breve{A}^\dagger_\mu$, $\lb N^l,
\hat{A}^\dagger_\mu\Breve{A}^\dagger_\mu \rb= \lb M^{rs},
\hat{A}^\dagger_\mu\Breve{A}^\dagger_\mu
\rb=0$. The dual vector field $\Breve{A}^\dagger_\mu$ transforms as in (\ref{v4}), with all $\Breve{A}_\mu$ standing to the furthest left  in any
expression replaced
by $\Breve{A}_\mu^\dagger$ standing to the furthest right.

\subsection{Vector fields related to frame one-forms}

In the same manner as we derived the vector fields corresponding to
$\hp_\mu$, we can also calculate vector fields
$\widetilde{A}_\mu$, corresponding to $\widetilde{\partial}_\mu$, the
derivative  dual   to the frame
one-forms up to the factor $\frac{1}{1-\frac{a^2}{4}\hat{\Box}}$. The calculation is much  simpler and we obtain:
\begin{eqnarray}
\label{v8}
\lb M^{rs}, \widetilde{A}_i\rb &=& \delta ^r_i \widetilde{A}_s -  \delta ^s_i \widetilde{A}_r,\qquad\qquad
\lb M^{rs}, \widetilde{A}_n\rb = 0, \nonumber\\
\lb N^l, \widetilde{A}_i\rb &=&- \delta ^l_i\sqrt{1-a^2\widetilde{\pat}_\mu\widetilde{\pat}_\mu}\widetilde{A}_n +ia\widetilde{\pat}_i \widetilde{A}_l-ia\de^l_i\widetilde{\pat}_\mu \widetilde{A}_\mu ,\\
\lb N^l, \widetilde{A}_n\rb &=&\Big(ia\widetilde{\pat}_n+\sqrt{1-a^2\widetilde{\pat}_\mu\widetilde{\pat}_\mu}\Big) \widetilde{A}_l. \nonumber
\end{eqnarray}
From (\ref{v8}) we could read off immediately the transformation
behaviour of $\widetilde{A}^\dagger_\mu$, but comparing with
$\widetilde{\Breve{A}}_\mu$, which can be obtained from the invariant
\begin{equation}
\label{v3a}
\lb M^{rs},\widetilde{\Breve{A}}_\mu\widetilde{A}_\mu\rb =0,\quad \textrm{ and  }\quad
\lb N^{l},\widetilde{\Breve{A}}_\mu\widetilde{A}_\mu\rb =0,
\end{equation}
we find that $\widetilde{A}^\dagger_\mu=\widetilde{\Breve{A}}_\mu$:
\begin{eqnarray}
\label{v7a}
\lb M^{rs}, \widetilde{A}^\dagger_i\rb &=& \delta ^r_i \widetilde{A}^\dagger_s
-\delta ^s_i \widetilde{A}^\dagger_r, \qquad\qquad
\lb M^{rs}, \widetilde{A}^\dagger_n\rb = 0, \nonumber\\
\lb N^l, \widetilde{A}^\dagger_i\rb &=&- \delta ^l_i\widetilde{A}^\dagger_n\sqrt{1-a^2\widetilde{\pat}_\mu\widetilde{\pat}_\mu} +ia \widetilde{A}^\dagger_l\widetilde{\pat}_i-ia\de^l_i\widetilde{A}^\dagger_\mu\widetilde{\pat}_\mu , \\
\lb N^l, \widetilde{A}^\dagger_n\rb &=& \widetilde{A}^\dagger_l\Big(ia\widetilde{\pat}_n+\sqrt{1-a^2\widetilde{\pat}_\mu\widetilde{\pat}_\mu}\Big) .\nonumber
\end{eqnarray}
In this sense, the vector field $\widetilde{A}_\mu$ is self-dual.

%
%
%
%
%
%

\subsection{Derivative-valued vector fields}

\vspace*{0.3cm}
We will now show that $\hat{A}_\mu$, $\Breve{A}_\mu$,
$\Breve{A}^\dagger_\mu$ and $\hat{A}^\dagger_\mu$ can be obtained from the vector field $\hat{V}_\mu$ by a
derivative-valued map
$\h{V}_\mu = \h{e}_{\mu\nu}\h{A}_\nu$,
$\h{A}_\mu = (\h{e}^{-1})_{\mu\nu} \h{V}_\nu$.
This is a change
in the basis of derivatives, $\h{e}_{\mu\nu}=\h{e}_{\mu\nu}(\partial)$ depends on the
derivatives.

We know the transformation properties of
$\hat{V}_\mu$, $\hat{A}_\mu$ and $\hat{\partial}_\mu$, (\ref{v1}),
(\ref{v2}) and (\ref{k12}). We expand these in powers
of $a$, at zeroth order we assume that $\h{V}_\mu\big|
_{\mathcal{O}(a^0)}=\h{A}_\mu\big| _{\mathcal{O}(a^0)}$ are the same vector field. We obtain a
recursion formula in $a$ that can be solved:
\begin{eqnarray}
\label{v9}
\h{e}_{nn} &=& {1\over a\hp _n}\sin (a\hp _n)+e^{-ia\hp _n}\Big( {ia\over 2}-{i\over \hp _n}\tan
\big( {a\hp _n\over 2}\big) \Big)
{\hp_k\hp_k \over \hp _n} , \nonumber\\
\h{e}_{nj} &=& {i\over \hp _n}e^{-ia\hp _n}\tan
\big( {a\hp _n\over 2}\big)\hp _j ,\nonumber\\
\h{e}_{ln} &=& \Big( e^{-ia\hp _n}-{1-e^{-ia\hp _n}\over ia\hp _n}\Big){\hp _l\over \hp _n} , \\
\h{e}_{lj} &=& {1-e^{-ia\hp _n}\over ia\hp _n}\delta _{lj}. \nonumber
\end{eqnarray}

To find the inverse of the matrix $\h{e}_{\mu\nu}$, we have to take care to single out the right partial derivatives. The result is:
\begin{eqnarray}
\label{v11}
(\h{e}^{-1})_{nn} &=& F^{-1}(\hat{\partial}_\mu) {e^{-ia\hp _n}-1 \over -ia\hp _n},\nonumber  \\
(\h{e}^{-1})_{nj} &=& F^{-1}(\hat{\partial}_\mu) \left(-{i\over \hp _n}
e^{-ia\hp _n} \tan({a\hp _n \over 2})\right) \hp_j,\nonumber\\
(\h{e}^{-1})_{ln} &=& F^{-1}(\hat{\partial}_\mu) \left({e^{-ia\hp _n}-1 \over -ia\hp _n}
-e^{-ia\hp _n}\right)\hp_l,
 \\
(\h{e}^{-1})_{lj} &=&  \frac{-ia\hp _n}{e^{-ia\hp _n}-1}\delta _{lj}
+\frac{{i\over \hp _n^2} e^{-ia\hp _n} \tan({a\hp _n \over 2})
\left(e^{-ia\hp _n}- {e^{-ia\hp _n}-1 \over -ia\hp _n} \right)
}{F(\hat{\partial}_\mu) \left(e^{-ia\hp _n}-1 \over -ia\hp _n \right)}\hp_l\hp_j,\nonumber\\
F(\hat{\partial}_\mu) &=&\left({1 \over ia^2 \hp _n ^2}\sin(a\hp _n)
\left ( 1-e^{-ia\hp _n} \right)-{ \hp_k\hp_k \over 2i\hp _n ^2} \tan({a\hp_n \over 2}) e^{-ia\hp _n}
\left( 1-e^{-ia\hp _n} \right) \right).\nonumber
\end{eqnarray}

The vector field $\Breve{A}_\mu$ is also defined by its transformation
behaviour that was derived from  (\ref{v3}). As
the derivatives are on the right of $\Breve{A}_\mu$ we make the ansatz
$\Breve{A}_\nu=\hat{V}_\mu(\Breve{e}^{-1}_{\mu\nu})$.

This ansatz is inserted into $\Breve{A}_\mu \h{A}_\mu=\hat{V}_\rho(\Breve{e}^{-1})_{\rho\mu}
(\h{e}^{-1})_{\mu\nu}\h{V}_\nu$.
But we know that $\h{V}_\mu\h{V}_\mu$ is an invariant, therefore
we conclude that $(\Breve{e}^{-1})_{\rho\mu}(\h{e}^{-1})_{\mu\nu}=\delta _{\rho\nu}$,
which leads to
\begin{equation}
\label{v13}
(\Breve{e}^{-1})_{\rho\mu}=\h{e}_{\rho\mu}.
\end{equation}
The formulae for $\Breve{A}^\dagger_\mu$ and $\hat{A}^\dagger_\mu$ are obtained
by conjugation.

There is also a transformation matrix $\widetilde{e}_{\mu\nu}$ from
 $\hat{V}_\mu$ to  $\widetilde{A}_\nu$
 (respectively $\widetilde{\Breve{A}}_\nu=\widetilde{A}^\dagger_\nu$) :
\begin{equation}
\label{v14}
\widetilde{A}_\mu  =  \widetilde{e}_{\mu\nu} \hat{V}_\nu,
\qquad \widetilde{A}^\dagger_\mu  = \hat{V}_\nu \widetilde{e}_{\mu\nu},
\end{equation}
which is
\begin{eqnarray}
\label{v15}
\widetilde{e}_{nn} &=& 1, \qquad\quad\hspace{2mm}
\widetilde{e}_{nj}=-ia\hdi_j\frac{ia\hdi_n +\sqrt{1-a^2\hdi_\mu\hdi_\mu}}{1-a^2\hdi_k\hdi_k}\nonumber\\
\widetilde{e}_{ln} &=& ia\hdi_l, \qquad
\widetilde{e}_{lj}=\de _{lj}\sqrt{1-a^2\hdi_\mu\hdi_\mu}+a^2\hdi_j\hdi_l\frac{ia\hdi_n +\sqrt{1-a^2\hdi_\mu\hdi_\mu}}{1-a^2\hdi_k\hdi_k}.
\end{eqnarray}
The inverse matrix $\hat{V}_\mu  = \widetilde{A}^\dagger_\nu
(\widetilde{e}^{-1})_{\mu\nu}$, with
$\widetilde{e}_{\nu\mu}(\widetilde{e}^{-1})_{\la\nu}=\de _{\mu\la}$ is
{\small \begin{eqnarray}
\label{v17}
(\widetilde{e}^{-1})_{nn} 
&=&1+\frac{a^2\widetilde{\pat}_k\widetilde{\pat}_k}{\sqrt{1-a^2\widetilde{\pat}_\mu\widetilde{\pat}_\mu}}\frac{-ia\widetilde{\pat}_n
  +\sqrt{1-a^2\widetilde{\pat}_\la\widetilde{\pat}_\la}}{1-a^2\widetilde{\pat}_s\widetilde{\pat}_s},\hspace{5mm}
(\widetilde{e}^{-1})_{nj}=\frac{ia\widetilde{\pat}_j}{\sqrt{1-a^2\widetilde{\pat}_\mu\widetilde{\pat}_\mu}},\nonumber\\
(\widetilde{e}^{-1})_{ln}&=&\frac{-ia\widetilde{\pat}_l}{\sqrt{1-a^2\widetilde{\pat}_\mu\widetilde{\pat}_\mu}}\frac{-ia\widetilde{\pat}_n
  +\sqrt{1-a^2\widetilde{\pat}_\la\widetilde{\pat}_\la}}{1-a^2\widetilde{\pat}_k\widetilde{\pat}_k},\hspace{10mm}
 (\widetilde{e}^{-1})_{lj}=\de _{lj}\frac{1}{\sqrt{1-a^2\widetilde{\pat}_\mu\widetilde{\pat}_\mu}}.
\end{eqnarray}}

\section{Conclusion}
In this work we have shown how to construct algebraic-geometric
quantities on
a specific noncommutative space, the $\kappa$-deformed space. This method allows to define algebraic-geometric quantities via their
consistency with the defining relations of the noncommutative space, adding a minimal set of
additional requirements. For example, we have shown how to construct
differential forms by demanding consistency with the
coordinate algebra and a specific transformation behaviour under $SO_a(n)$ rotations.

The method presented here does not require a thorough
understanding of deep mathematical concepts such as Hopf
algebras, for treating noncommutative spaces. The Hopf algebra description of the $SO_a(n)$ symmetry
however can be fully recaptured in this context. 

For more general noncommutative spaces than the $\ka$-deformed space, our approach might not automatically
lead  to well-founded results. However, since this method has shown so fruitful
here, leading directly to workable definitions of derivatives,
differential forms and 
vector fields, we suggest that the presented method can be used also
to investigate other noncommutative spaces.

\section*{Acknowledgements}
We are indebted to Julius Wess for many ideas and inspirations,
strongly influencing the work on 
 this article. Also we are grateful to Larisa Jonke for fruitful discussions and careful
proof-reading of this article.


\end{document}